\documentclass{nobanner}

\usepackage[english]{babel}
\usepackage[T1]{fontenc}
\usepackage{graphicx}
\usepackage{newtxtext}
\usepackage{newtxmath}
\usepackage{comment}
\usepackage{amsfonts,mathtools,amssymb,stmaryrd}
\usepackage{color,xcolor}
\usepackage[breaklinks,colorlinks=true,allcolors=blue]{hyperref}
\usepackage{epstopdf, epsfig}
\usepackage{subfig}
\usepackage{natbib}
\usepackage{grffile}
\usepackage{amsmath}
\usepackage{esint}

\newcommand{\mH}{{\mathcal H}}
\newcommand{\mE}{{\mathcal E}}
\newcommand{\mI}{{\mathcal I}}
\newcommand{\mf}{{\mathrm{f}}}

\newcommand{\bl}{{\boldsymbol \ell}}

\newcommand{\ba}{{\ensuremath\boldsymbol{a}}}

\newcommand{\bx}{{\ensuremath\boldsymbol{x}}}
\newcommand{\bu}{{\ensuremath\boldsymbol{u}}}

\newcommand{\br}{{\ensuremath\boldsymbol{r}}}
\newcommand{\bk}{{\ensuremath\boldsymbol{k}}}

\newcommand{\be}{\begin{equation}}
\newcommand{\ee}{\end{equation}}

\newcommand{\hit}{SQG-HIT}

\newcommand{\av}[1]{\left\langle{#1}\right\rangle}

\newcommand{\cb}{\color{black}}

\newcommand{\inria}{Universit\'e C\^ote d'Azur, Inria, CNRS, Calisto team, 06902 Sophia Antipolis, France}
\newcommand{\inphyni}{Universit\'e C\^ote d'Azur, CNRS, Institut de Physique de Nice, 06200 Nice, France}
\title{Surface quasigeostrophic turbulence: The refined study of an active scalar}
\author{Nicolas Valade\aff{1}\corresp{\email{nicolas.valade@inria.fr}}, %
J\'er\'emie Bec\aff{1,2}\corresp{\email{jeremie.bec@univ-cotedazur.fr}}, %
and Simon Thalabard\aff{2}\corresp{\email{simon.thalabard@univ-cotedazur.fr}}
}
\affiliation{\aff{1} \inria\\ \aff{2} \inphyni} 
\date{\today}

\graphicspath{{./Images/I/}{./Images/II/}{./Images/III/}{./Images/IV/}{./Images/}{./}}
\begin{document}
\maketitle

\begin{abstract}
    SQG describes the two-dimensional active transport of a scalar field, such as temperature, which---when properly rescaled---shares the same physical dimension of length/time as the advecting velocity field.  This duality has motivated analogies with fully developed three-dimensional turbulence. In particular, the Kraichnan-Leith-Batchelor similarity theory predicts a Kolmogorov-type inertial range scaling $\propto (\varepsilon \ell)^{1/3}$ for both scalar and velocity fields, and the presence of intermittency through multifractal scaling was pointed out by Sukhatme \& Pierrehumbert (\textit{Chaos}~\textbf{12}, 439, 2002) in  unforced settings.
In this work, we refine the discussion of these statistical analogies, using numerical simulations with up to $16,384^2$ collocation points in a steady-state regime dominated by the direct cascade of scalar variance. 
We show that  mixed structure functions, coupling velocity increments with scalar differences, develop well-defined scaling ranges, highlighting the role of anomalous fluxes  of all the  scalar moments.
However, the clean multiscaling properties of SQG transport are blurred  when considering velocity and scalar fields separately. 
In particular, the  usual (unmixed) structure functions  do no follow any power-law scaling in any  range of scales, neither for the velocity nor for the scalar increments.  
This specific form of the intermittency phenomenon reflects the specific kinematic properties of SQG turbulence, involving the interplay between  long-range interactions, structures and geometry.  Revealing the multiscaling in  single-field statistics requires to resort to generalised notions of scale invariance, such as extended self-similarity and specific form of refined self-similarity.
Our findings emphasise the fundamental entanglement of scalar and velocity fields in SQG turbulence: They evolve hand in hand and any attempt to isolate them destroys scaling in its usual sense. This perspective sheds new lights on the discrepancies in spectra and structure functions, that have been repeatedly observed in SQG numerics for the past 20 years.
\end{abstract}        

\section{Introduction}
\label{section1}
The surface quasi-geostrophic (SQG) theory describes the dynamics of rapidly rotating and stably stratified flows near flat horizontal boundaries through the two-dimensional (2D) transport equation
\be
\label{eq:sqg}
\begin{split}
	 & \partial_t \theta + \left(\bu \bcdot \bnabla\right)\theta = \nu \Delta \theta + F, \quad \bu  := -\boldsymbol{\nabla}^\perp \psi, \quad \psi:=(-\rmDelta)^{-\frac{1}{2}}\theta,
\end{split}
\ee
where $\theta(\bx,t)$ represents a scalar field (such as surface temperature or buoyancy) advected in a planar domain $\rmOmega\subset\mathbb{R}^2$ under the influence of forcing $F(\theta, \bx,t)$ and viscous damping. Here, $\bnabla^\perp = (-\partial_{2},\partial_{1})^\top$ denotes the perpendicular gradient operator. 
Initially proposed by \cite{blumen1978uniform}, the SQG model has emerged as a minimal yet insightful framework for studying the dynamics of upper-oceanic and lower-tropospheric flows. It is particularly relevant to the problems involving potential vorticity inversion and frontogenesis \citep[see, for example,][and the comprehensive review by \citealt{lapeyre2017surface}]{juckes1995instability,held1995surface,lapeyre2006dynamics}.  At a more fundamental level, the SQG system \eqref{eq:sqg} has also gathered significant interest in physical and mathematical turbulence research due to its phenomenological and structural analogies to various prototypical fluid systems. These include the 2D Navier--Stokes equations \citep{held1995surface,bernard2007inverse}, passive scalar turbulence \citep{celani2004active,llewellyn2016interacting}, Burgers  turbulence \citep{valade2024anomalous}, and the 3D Navier-Stokes equations \citep{constantin1994formation,sukhatme2002surface}.

\begin{figure}
\includegraphics[width=.296\textwidth,trim=0cm 0cm 0cm 0cm, clip]{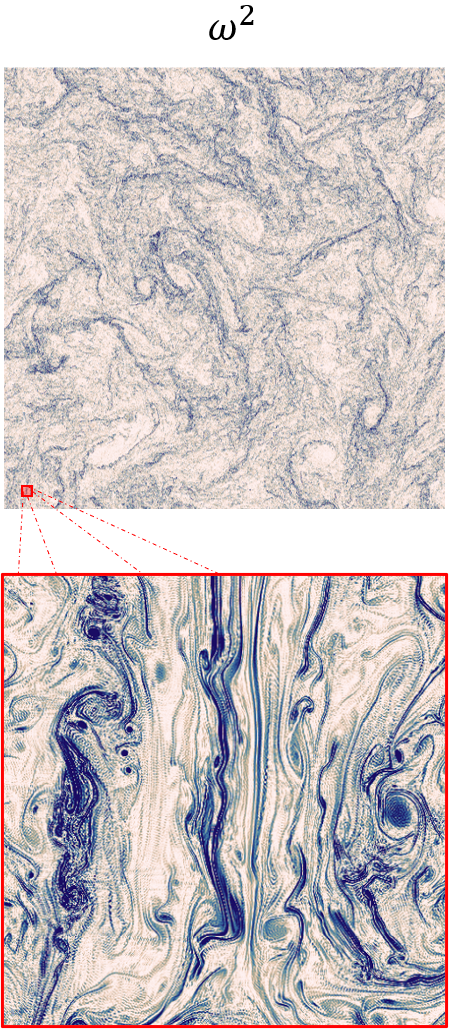}
\includegraphics[width=.299\textwidth,trim=0cm 0cm 0cm 0cm, clip]{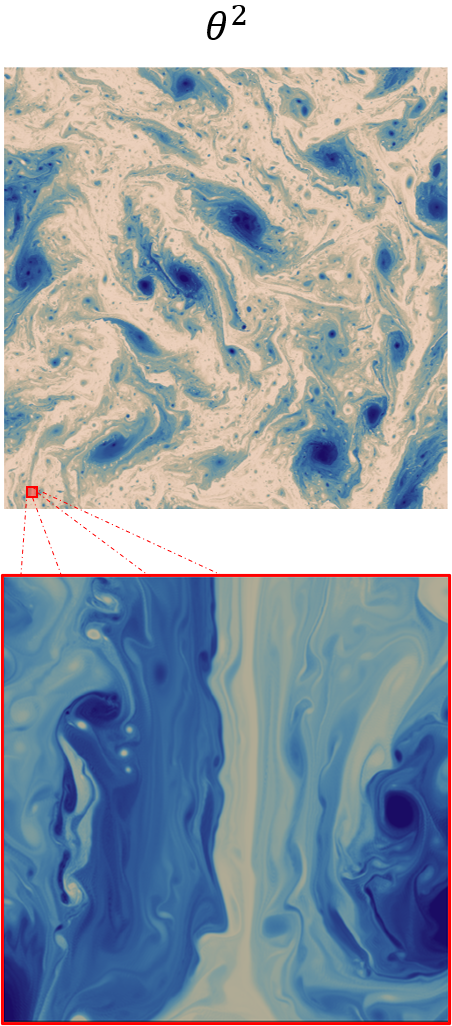}
\includegraphics[width=.30\textwidth,trim=0cm 0cm 0cm 0cm, clip]{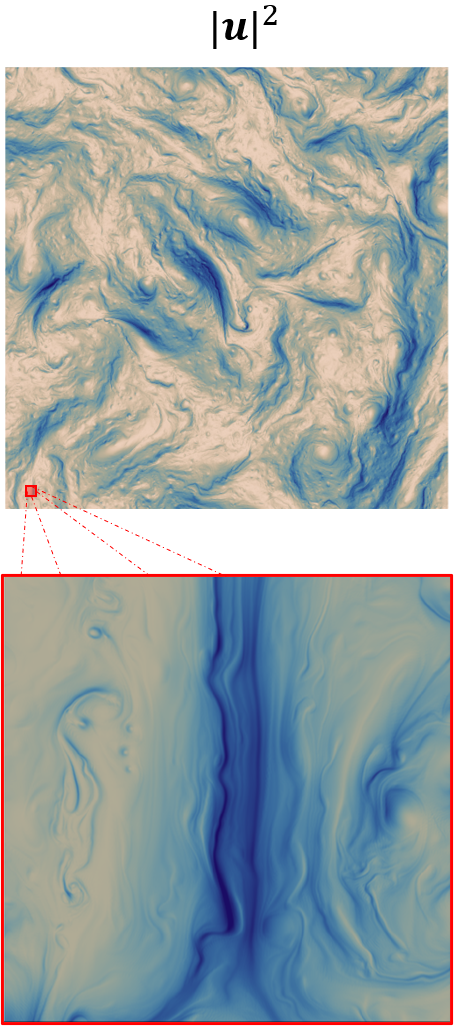}
\includegraphics[width=.078\textwidth,trim=0cm 0cm 0cm 0cm, clip]{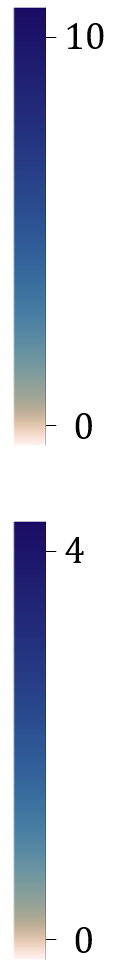}

\caption{ SQG snapshots of squared  vorticity $\omega=\bnabla^\perp \bcdot \bu$, temperature $\theta$ and  velocity  $\bu$ at a typical time. The zoomed-in regions display $(1/32)^2$ of the computational domain. Taken from Run VI (see Tab.~\ref{Tab1:Settings}). 
}
\label{fig:1}
\end{figure}

Equation \eqref{eq:sqg} describes the transport of an \textit{active scalar}, where the velocity field $\bu$ depends nonlocally on the advected scalar $\theta$, analogous to the velocity--vorticity relationship in the 2D Euler equations. In SQG, though, long-range interactions are mediated by  the Green function $(-\Delta)^{-1/2}$ associated to the half-Laplacian rather than to the Laplacian. In the full plane, this yields 
\be \label{NLOperators} 
\psi(\bx) =\frac{1}{2\pi} \int_{\mathbb R^2} \frac{\theta(\bx')}{|\bx-\bx'|}\,\mathrm{d}^2\bx'\quad \text{ and } \quad
\bu(\bx) =-\frac{1}{2\pi} \int_{\mathbb R^2} \frac{ (\bx-\bx')^\perp}{|\bx-\bx'|^3} \theta(\bx')\,\mathrm{d}^2\bx',
\ee
where $\bx^\perp = (-x_2,x_1)$. 
These relations imply, somewhat surprisingly, that $\theta$ and $\bu$ share the same physical dimension. This makes $\theta$ comparable to both a coarse-grained vorticity and a scalar analog of the velocity field (see Fig.~\ref{fig:1}).  In the absence of forcing and dissipation, smooth solutions of the SQG equation conserve two quadratic invariants: the Hamiltonian $\mathcal{H}= {\frac{1}{2}\int_\Omega \psi\,\theta}$ %
akin to the energy in 2D Euler flows, and the scalar variance, which also corresponds to the kinetic energy,
$\mathcal{E}= \frac{1}{2} \int_\Omega \theta^2=\frac{1}{2} \int_\Omega |\bu|^2$ %
akin to the enstrophy. 
In the presence of forcing and dissipation, these invariants lead to a dual cascade scenario, featuring an inverse cascade of the Hamiltonian and a direct cascade of scalar variance  \citep{burgess2015kraichnan, smith2002turbulent}.

The direct transfer of scalar variance---and equivalently of kinetic energy---is known to differ fundamentally from the enstrophy cascade in 2D turbulence, as it sustains more energetic small scales.  In short, SQG turbulence is rough.  Similar to 3D turbulence, dimensional arguments tie the direct SQG cascade to a Kolmogorov energy spectrum $\propto k^{-5/3}$. However, previous numerical studies have reported both steeper spectral slopes \citep{sukhatme2002surface,celani2004active, capet2008surface} and shallower ones \citep{watanabe2007interacting}. 
These observations point toward the presence of significant corrections due to intermittency \citep{kolmogorov1962refinement,frisch1995turbulence,benzi2023lectures}. One could also question the extent to which  SQG scaling laws (if any) are universal.
In particular, the recent work of \cite{valadao2024non} provided evidence that large-scale fluctuations can obscure a $k^{-5/3}$ scaling, with steeper subleading terms appearing in the spectrum.

In this work, we aim to examine in greater detail the Kolmogorovian features of steady-state SQG turbulence through a series of highly resolved direct numerical simulations, employing up to $16,384^2$ collocation points. These simulations incorporate small-scale dissipation and a statistically homogeneous large-scale forcing scheme to sustain turbulence. Classical 3D turbulence theory relies on the phenomenological assumptions that the steady-state velocity fields have finite variance and that the kinetic energy dissipation rate remains nonzero in the limit of vanishing viscosity---a property referred to as anomalous dissipation. For SQG turbulence, these conditions translate into the requirements:
\be
\label{eq:anomalous}
 \lim_{\nu\to 0} \av{\theta^2}< \infty \quad \text{and } \quad \lim_{\nu\to 0} \av{\nu|\bnabla \theta|^2 }>0,
\ee
where $\av{\cdot}$ denotes a suitable space-time averaging. We propose a numerical setup designed to check whether these conditions are satisfied. %
We point out that their realisation in SQG turbulence is not straightforward: The presence of Hamiltonian inverse transfers requires  large-scale damping  to establish a statistically steady state. Since interactions between scales in SQG might be nonlocal  \citep[see, e.g.][]{watanabe2007interacting,foussard2017relative}, this  damping could influence the statistics of the direct cascade.

Our numerical simulations enable a refined analysis of the multiscaling properties of SQG.
In 3D turbulence, multiscaling is well captured by \citealt{kolmogorov1962refinement} refined similarity hypothesis, which links the non-Gaussianity of velocity increments to fluctuations in the dissipation field \citep{chevillard2019skewed,dubrulle2019beyond}. 
While evidence of intermittency in SQG turbulence has been reported in unforced settings  \citep{sukhatme2002surface,valade2024anomalous},
its phenomenology remains poorly understood. Qualitatively, coherent structures play a pivotal role in SQG dynamics across scales, generating sharp scalar gradients and complex velocity patterns entangling filaments, fronts, and vortices, as illustrated in Fig.~\ref{fig:1}.  
These structures are likely coupled through cascades of shear instabilities \citep{scott2014numerical,lapeyre2017surface}. As a result, although the scalar field $\theta$ and the velocity field $\bu$ share certain features---such as their variance, dissipation rate, and dimensional scaling $\propto (\varepsilon \ell) ^{1/3}$---they exhibit distinct statistical behaviours. For instance, \citet{capet2008surface} observed that the longitudinal velocity increment $\delta u_\parallel(\ell)$ has a positive skewness, while $\delta \theta$ is symmetric. Moreover, the direct cascade of scalar variance imposes  a negative correlation between $\delta u_\parallel$ and $\left(\delta\theta\right)^2$, stemming from a SQG-specific version of Yaglom's law \citep{yaglom1949local}.
As a specificity of SQG turbulence, these structural differences raise questions about its classification relative to 2D and 3D and about the quantification of intermittency. In particular, distinct scaling laws, if they exist, may apply to the scalar and velocity fields individually.

To address these questions, the paper is organised as follows.  \S\ref{section2} introduces our numerical setup and discusses the establishment of a statistically steady SQG state in a periodic domain.  \S\ref{section3} investigates inertial-range scaling, with a particular focus on Yaglom's law and its implications for the statistical properties of both velocity and scalar fields. \S\ref{section4} explores intermittency by analyzing higher-order structure functions and scalar dissipation statistics within the framework of the refined self-similarity hypothesis.  Finally, \S\ref{section5} summarises  our findings and formulates concluding remarks.
Some technical but otherwise standard aspects of SQG are compiled in Appendix.
\section{Steady-state SQG turbulence in a doubly-periodic domain}
\label{section2}

\subsection{SQG turbulence}
\label{ssec:SQGturb}
\subsubsection{Steady-state balances}
Our  numerical simulations implement the SQG system \eqref{eq:sqg} in the doubly $2\pi$-periodic domain, $\rmOmega=\mathbb{T}^2$, by decomposing the fields into their Fourier components. 
The Fourier transform of the scalar field is defined as $\hat{\theta}_\bk = \frac{1}{(2\pi)^2} \int_{\mathbb{T}^2} \theta(\bx)\,\mathrm{e}^{-\mathrm{i}\,\bk\bcdot\bx}\,\mathrm{d}^2x$, such that $\theta(\bx) = \sum_{\bk\in\mathbb{Z}^2} \hat{\theta}_\bk \mathrm{e}^{\mathrm{i}\,\bk\bcdot\bx}$, and similarly for the components of the velocity field.
In Fourier space, the velocity-scalar SQG relationship (Riesz transform) is then mediated through the explicit identities
\begin{equation} \label{eq:riesz_fourier}
    \hat{\bu}_\bk = -\mathrm{i}\frac{\bk^\perp}{|\bk|}\hat{\theta}_\bk \quad  \text{and} \quad   \hat{\theta}_\bk = \mathrm{i}\frac{\bk^\perp }{|\bk|} \bcdot \hat{\bu}_\bk \quad\text{for }  \bk\in\mathbb{Z}^2\backslash\{\boldsymbol{0}\},
\end{equation}
where the null components are set to zero:  $\hat{\bu}_{\boldsymbol{0}} = \boldsymbol{0}$ and $\hat \theta_{\boldsymbol{0}} = 0$.
From  \eqref{eq:riesz_fourier}, it follows that the Fourier components of $\theta$ and $\bu$ have the same amplitudes: $|\hat{\bu}_\bk| = |\hat{\theta}_\bk|$ for all $\bk$. As a result, they share the same power spectrum 
$$E(k) = \dfrac{1}{2}\!\!\!\!\!\!\sum_{~~~k\le|\bk|<k+1}\!\!\!\!\!\!\!\!\!\!|\hat{\theta}_{\boldsymbol{k}}|^2 = \dfrac{1}{2}\!\!\!\!\!\!\sum_{~~~k\le|\bk|<k+1} \!\!\!\!\!\!\!\!\!\! |\hat{\bu}_\bk|^2;$$
as well as their spatially-averaged square values and gradient norms
\be
	\fint\theta^2 = \fint |\bu|^2 \quad \text{and} \quad \fint |\bnabla \theta|^2 = \fint |\bnabla \bu |^2,
\ee
where we adopt the shorthand $\fint =  \int_{\mathbb{T}^2} /(2\pi)^2$ for the spatial average over  $\mathbb{T}^2$.
To achieve a statistically steady state, we employ a stochastic forcing scheme $F(\theta,\bx,t)$ that combines the superposition of white-in-time Gaussian random Fourier modes with a linear damping operating solely at the largest scales. In Fourier space, the forcing takes the form
\be
         \label{eq:forcing}
	\begin{split}
	 & \hat F_\bk = \sqrt{\dfrac{2 \mathcal I}{c_d\pi k_\mf^2}}\,\mathrm{e}^{-\frac{|\bk|^2}{2k_\mf^2}}\,\hat\eta_{\bk} - \alpha \,\hat \theta_{\bk}\, {\mathbf 1}_{|\bk| \le k_\mf}. \\
	\end{split}
\ee
Here, $\eta_{\bk}$ are random complex Gaussian coefficients, with statistics prescribed by $\mathbb{E}\left[\hat{\eta}_{\bk}(t)\right] = 0$ and $\mathbb{E}\left[\hat{\eta}_{\bk}(t)\hat{\eta}_{\bk'}(t')\right]=\delta_{\bk+\bk'}\,\delta(t-t')$, where $\mathbb{E}[\cdot]$ denotes averaging over random realisations.
The numerical factor $c_d \simeq 1.004$ accounts for discretisation effects and is such that the parameter $\mathcal I$ represents the injection rate of energy. The damping coefficient $\alpha$ regulates the magnitude of large-scale contributions to the energy and the Hamiltonian. In the stationary state,
by letting $\theta_\mf = \sum_{|\bk| \le k_\mf} \hat\theta_\bk\,\mathrm{e}^{\mathrm{i}\,\bk \bcdot \bx}$ and 
$\psi_\mf = \sum_{|\bk| \le k_\mf} |\bk|^{-1}  \hat\theta_\bk\,\mathrm{e}^{\mathrm{i}\,\bk \bcdot \bx}$, the following global balances for the  energy and the  Hamiltonian hold:
\be
	\label{eq:global_balances}
	\mathcal I = \varepsilon +\alpha\, {\av{\theta^2_\mf}}  \quad \text{ and }\quad  k^{-1}_\mI \mI= \gamma + \alpha\, {\av{\theta_\mf\, \psi_\mf}},
\ee
where $\varepsilon := \nu {\av{|\bnabla \theta|^2}}$,  $\gamma := \nu {\av{ \bnabla \theta \bcdot \bnabla \psi}}$, and the angular brackets $\av{\cdot}$ denote averages over space, time, and forcing realisations. In the Hamiltonian balance, we introduced the centroid wavenumber $k_\mI = \tilde c_d k_\mf/\sqrt \pi$ (with $\tilde c_d \simeq 1.301$), which defines an injection scale.

\subsubsection{Definition of SQG turbulence}
\label{ssc:SQGturb}
Following the review by \cite{boffetta2012two} on the 2D Navier--Stokes equations,  we introduce the characteristic SQG damping and viscous centroid wavenumbers as, respectively,
\be
\label{eq:scales}
 	k_\alpha=  \dfrac{\av{\theta_\mf^2}}{\av{\theta_\mf \,\psi_\mf}} \quad \text{and} \quad k_\nu =  \dfrac{\varepsilon}{\gamma}.
\ee
On phenomenological grounds, we expect $k_\nu>k_\mI>k_\alpha$, with the magnitudes of the scale separations defining  steady-state regimes with different physical properties.
In this work, we \emph{define} SQG turbulence as a steady-state regime satisfying 
\be
	\label{eq:SQGturb}
	k_\nu \gg k_\mI \gtrsim k_\alpha, \text{ with } k_I,k_\alpha = O(1).
\ee
 Under this assumption, the steady-state balances~\eqref{eq:global_balances}, combined with the definitions \eqref{eq:scales}, yield the following asymptotic estimates
\be
	\label{eq:cond}
	  \varepsilon \simeq \left(1-\dfrac{k_\alpha}{k_\mI}\right)\mI,\quad
	{\av{\theta_\mf^2}} \simeq \dfrac{k_\alpha}{\alpha k_\mI} \mI, \quad
	 \text{and }  \gamma \simeq \dfrac{\varepsilon}{k_\nu}.
\ee

The estimates \eqref{eq:cond} fulfill the Kolmogorovian requirements \eqref{eq:anomalous} of anomalous scalar dissipation and ensure the finiteness of the scalar variance. Anticipating on the K41 phenomenology discussed in \S\ref{ssec:K41}, we expect the SQG spectra to have finite ultraviolet capacity, \textit{i.e.}, $\sum_{k >k_\mf} E(k) <\infty$. Equation~\eqref{eq:cond} also implies that the viscous dissipation of the Hamiltonian vanishes in the limit $\nu \to 0$. As such, our definition of SQG turbulence through Eq.~\eqref{eq:SQGturb}  formalises a steady-state regime sustaining a direct cascade  of scalar variance, with no inverse cascade.
At this stage, SQG turbulence already differs from 3D turbulence, as the energy dissipation rate does not perfectly balance the injected energy. The difference $\simeq\alpha\av{\theta_\mathrm{f}^2} = (k_\alpha/k_\mI) \mI$ corresponds to energy leaking into the largest scales.
This leakage introduces a degree of non-universality, but it comes as a necessary trade-off for the requirement that both $\alpha$ and $k_\alpha$ be finite, and prevents the scalar variance from blowing up.

\subsubsection{A comment on SQG scales}\label{ssec:sqgscales}
From the SQG turbulence asymptotics \eqref{eq:cond}, it is natural to define the SQG turnover time at wavenumber $k$ as $\tau := \varepsilon^{-1/3}k^{-2/3}$. We therefore define the timescales of injection $\tau_\mI$ and dissipation $\tau_\nu$ as the turnover times at wavenumbers $k_\mI$ and $k_\nu$, respectively. 
The dissipative scales $k_\nu$ and $\tau_\nu$ \emph{a priori} differ from the usual Kolmogorov scales $k_\eta = (\varepsilon/\nu^3)^{1/4}$, $\tau_\eta  = (\nu/\varepsilon)^{1/2}$ found in 3D turbulence. We postpone further comments on this issue to \S\ref{ssec:convergence}. We denote $\tau_\alpha=1/\alpha$ as the damping timescale. SQG turbulence should correspond to a situation where $\tau_\alpha \gg \tau_\mI$, in order to  prevent large-scale fluctuations from dominating the turbulent statistics of the direct cascade \citep[see][for a related discussion on the 3D Navier--Stokes equations]{homann2013effect}. The conversion between physical lengthscales and Fourier wavenumbers is mediated through the convention $\ell = 2\pi / k$.

\subsection{Numerical setup} 
\begin{table}
  \begin{center}
  \def~{\hphantom{0}}
  \begin{tabular}{crrcccccc}
    \textbf{Runs} & $N^2$~&$\nu$~~~~& $\Rey$ & $k_\mathrm{max}/k_\eta $ & $k_\mathrm{max}/k_\nu$ & $N_\mathrm{real}$ & $T_\mathrm{max}/\tau_\mI $ & $\av{\theta_\mf^2}$ \\\hline
    \textbf{I}    & $1\,024^2$   & $3.0\times10^{-4}$    & $2.3 \times 10^4$&1.0& 28 & 1 & $30$& 7.4\\
    \textbf{II}   & $2\,048^2$   & $1.2\times10^{-4}$ & $5.8 \times 10^4$& 1.0& 30 & 1 & $30$& 6.8\\
    \textbf{III}  & $4\,096^2$   & $4.7\times10^{-5}$ & $1.5 \times 10^5$& 1.0& 33 & 1 & $300$& 7.0\\
    \textbf{IV}  & $8\,192^2$   & $1.9\times10^{-5}$ & $3.7 \times 10^5$& 1.0& 35 & 4 & $7$& 7.3\\
    \textbf{V}   & $16\,384^2$ & $7.5\times10^{-6}$ & $9.0 \times 10^5$& 1.0& 36 & 4 & $7$& 6.8\\
    \textbf{VI}  & $16\,384^2$ & $2.0\times10^{-6}$    & $3.5 \times 10^6$& 0.4& 7 & 4 & $7$& 6.7
  \end{tabular}
  \caption{\textit{Numerical settings.}  See the text for definitions. In all runs, we set the damping rate to $\alpha=0.1$ and prescribe $k_\mf=3$ and $\mI=1$, corresponding to an injection wavenumber $k_\mI \simeq 2.2$.}
  \label{Tab1:Settings}
  \end{center}
\end{table}

Our numerical simulations of the forced-dissipated SQG equation~\eqref{eq:sqg} in the doubly $2\pi$-periodic torus $\mathbb{T}^2$ employ a GPU-based pseudo-spectral solver. The solver uses the standard $2/3$-rule for dealiasing \citep{orszag1971elimination}, fourth-order Runge--Kutta time marching, and up to $N^2=16,384^2$  collocation points. 
The forcing scheme \eqref{eq:forcing} uses $k_\mf = 3$, an injection rate $\mI=1$, and a damping rate $\alpha=0.1$, yielding an injection scale of $\ell_\mI = 2\pi/k_\mI \simeq 2.8$. To address finite-viscosity effects, we define the Reynolds number
\be
	  \Rey := \dfrac{\ell_\mI \av{\theta_\mathrm{f}^2}^{1/2}}{\nu}.
\ee 
Table~\ref{Tab1:Settings} summarises the key parameters of our simulation series. Across the six runs, the injection and damping parameters remain fixed, while the Reynolds number spans over two decades.
It is customary to characterise infrared damping in terms of a damped Reynolds number. Here, we define $\Rey_\alpha := \av{\theta_\mathrm{f}^2}^{1/2}/(\alpha \ell_\mI) \simeq 10$, which remains constant across all simulations. The damping rate $\alpha = 0.1$ to which this corresponds was determined through trial and error, to ensure that large-scale dynamics evolve on timescales significantly slower than the injection timescale, specifically  $\tau_\alpha\simeq 10  \simeq 4 \tau_\mI$, in line with \S\ref{ssec:sqgscales}. 

For Runs I to V, the maximal wavenumber $k_\mathrm{max} = N/3$ is chosen to satisfy $k_\mathrm{max} \simeq \,k_\eta$, following good practices in numerical studies of 3D turbulence \citep{grauer2012longitudinal,iyer2015refined,buaria2019extreme}.  This choice is relatively conservative in the context of SQG turbulence. As noted in \S\ref{ssec:sqgscales}, the relevant dissipative scale for SQG is $k_\nu$, which is amply resolved for these runs. 
This motivates the parameters of  Run~VI, trading a  higher Reynolds number for a less conservative scheme with $k_\mathrm{max} \simeq 0.4 k_\eta$.

Regarding simulation durations, Runs~I and II consist of a single realisation over $T_\mathrm{max} = 10\,\alpha^{-1} \simeq 30 \,\tau_\mI$. Runs~IV to VI gather statistics from four realisations, each initialised from different equilibrated states and subjected to distinct forcing realisations, with $T_\mathrm{max} = 2.5\,\alpha^{-1} \simeq 7 \,\tau_\mI$. Run~III corresponds to a single, much longer realisation, specifically designed to evaluate statistical convergence, as detailed in \S\ref{ssec:convergence}.

\subsection{Convergence towards SQG turbulence}
\label{ssec:convergence}
We here assess the convergence of our numerical protocol towards SQG turbulence, by verifying that the conditions in equations~ \eqref{eq:SQGturb} and \eqref{eq:cond} are satisfied as  $\Rey$ increases.

\subsubsection{Steady-state and stationary statistics}
\begin{figure}
\includegraphics[width=1\linewidth]{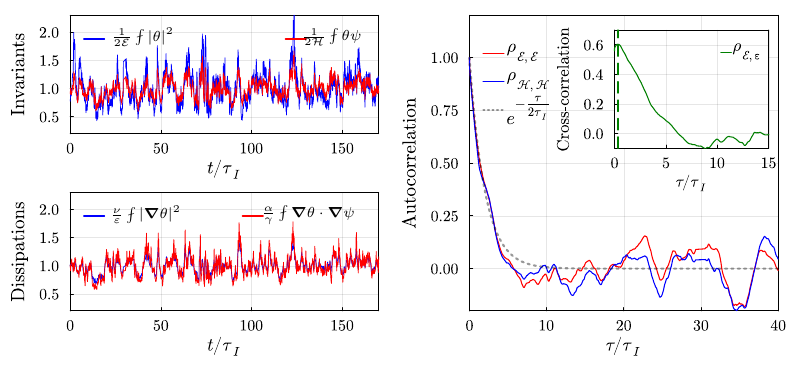}
\caption{ \emph{ Left:}  Time series of the SQG invariants and their corresponding dissipation rates for Run III, normalised by their time-averaged values. Right: Temporal correlations of scalar variance, Hamiltonian and cross-correlation of the energy and its dissipation rate.}
\label{fig:2}
\end{figure}
To evaluate the relevance of our numerical protocol to reach a statistical steady state, we first focus on Run III,  which features the longest time series, extending up to $300 \tau_\mI$.
Figure~\ref{fig:2} displays the typical temporal evolution of  the energy $\frac{1}{2}\fint \theta^2$ and the Hamiltonian $\fint \theta  \psi$, as well as their instantaneous dissipation rates
$\nu \fint |\bnabla  \theta|^2$ and $\nu \fint \bnabla  \theta \bcdot \bnabla  \psi$, normalised by their time average values $\mE$, $\mH$, $\varepsilon$, and $\gamma$, respectively.
It shows that all time series are statistically stationary, with well defined averages  and correlation-time $\tau_c\sim 2 \tau_\mI$.
With  $\tau_I$ emerging as the characteristic correlation time, it is justified to consider averages over durations of the order of tens of $\tau_\mI$ as representative of long-time steady-state statistics.

The statistics also reveal a high degree of cross-correlations.
The apparent synchronisation between the energy and the Hamiltonian reflect  the global balance between  injection and large-scale damping. In turn, the cross-correlation shown between the dissipation rate $\nu \fint |\bnabla  \theta|^2$ and the energy $\frac{1}{2}\fint \theta^2$ (Inset of the right panel) reflects the   direct cascade mechanism. The cross-correlation peaks at  $\tau \simeq 0.3\tau_\mI$ corresponds to the typical cascading time from injection downto dissipation scales.

\subsubsection{Scalar variance and dissipation}
\begin{figure}
\includegraphics[width=0.99\linewidth]{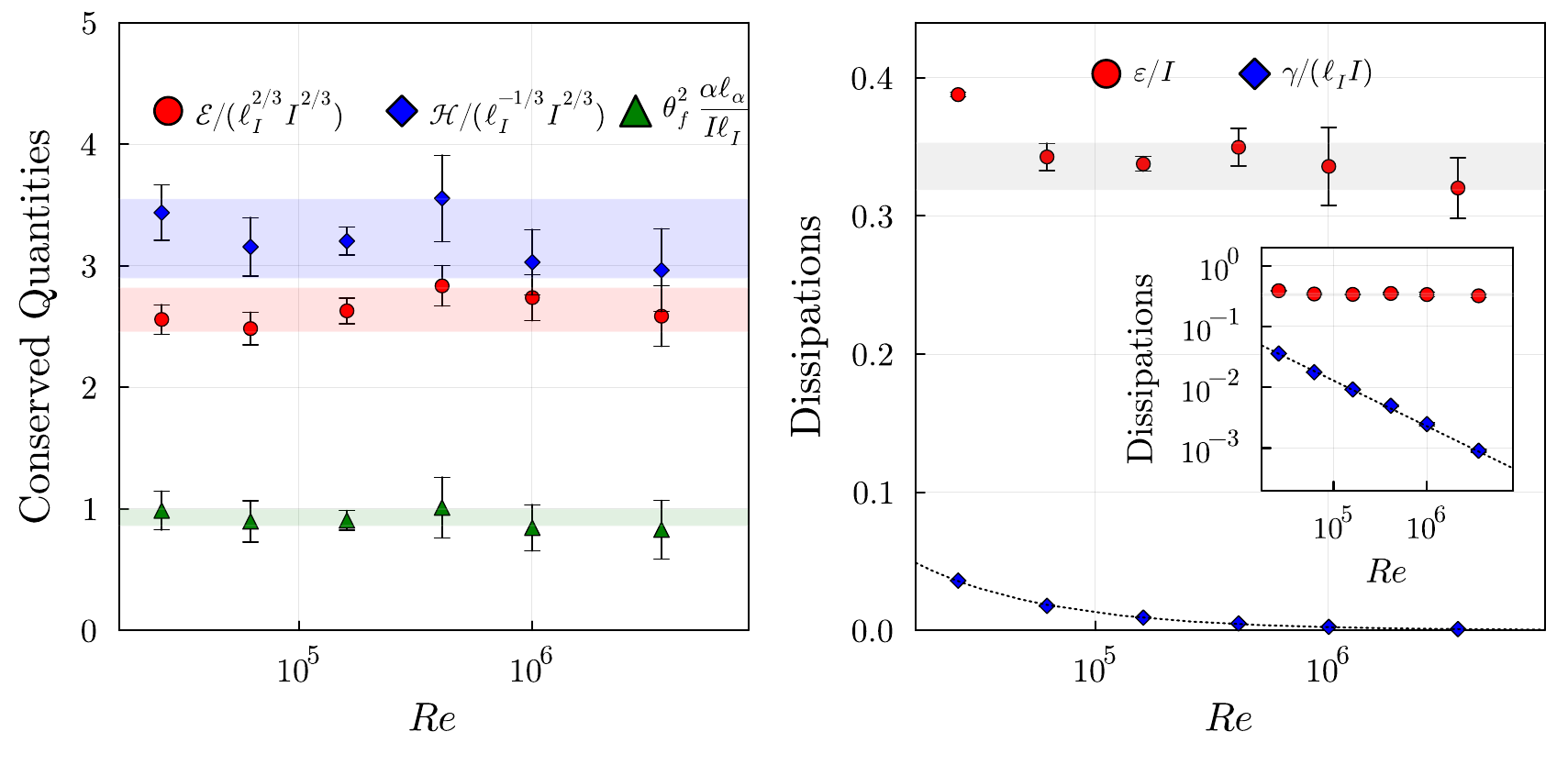}
\caption{\emph{Left:} Averaged values of the Hamiltonian $\mH$ and energy $\mE$ as a function of $\Rey$. \emph{Right:} Corresponding dissipation rates, $\gamma$ and $\varepsilon$, respectively. All these quantities are here normalised by K41 dimensional predictions incorporating the appropriate powers of  $\mI$ and $\ell_\mI$. }
\label{fig:3}
\end{figure}

The convergence of SQG dynamics towards a developed turbulent state is examined in figure~\ref{fig:3}, which displays the steady-state values of the quadratic SQG invariants and their dissipation rates as a function of the Reynolds number $\Rey$. Despite significant statistical fluctuations, our numerical results suggest 
that the energy  $\mE$ and the Hamiltonian $\mH$ converge towards asymptotic values approximately given by 
\be
	\mE \; \simeq 2.7\;  \ell_\mI^{2/3} \mI^{2/3},\quad \ell_\mI \mH \; \simeq 3.2\; \ell_\mI^{2/3} \mI^{2/3}
\ee
The distinct roles played by the energy and the Hamiltonian dissipation rates in establishing this asymptotic turbulent regime are evident from their respective dependencies on $\Rey$. While $\gamma$  algebraically vanishes as $\Rey$ increases, $\varepsilon$ asymptotes to a finite value:
\be
	\label{eq:anomalousdissip}
	\varepsilon \simeq 0.34 \mI,\quad \gamma \propto \Rey^{-3/4}.
\ee
The $-3/4$ scaling for the Hamiltonian dissipation is consistent with basic phenomenological arguments (discussed below). For the energy dissipation, the observed asymptotic value of $\varepsilon$ implies, together with the steady-state balance \eqref{eq:cond}, that $k_\alpha \simeq 0.65 k_\mI$. This indicates that more  than  60\% of the injected energy remains at the large scales and is damped by the $\alpha$ term. 
It is worth noting that a slow (for instance logarithmic) decay of $\varepsilon_\nu$ as a function of $\Rey_\nu$ cannot be entirely ruled out: The high level of statistical fluctuations in the dissipation field at the highest Reynolds numbers produce high uncertainties on the average estimates.

\subsubsection{Higher-order moments}
\begin{figure}
\includegraphics[width=0.99\linewidth]{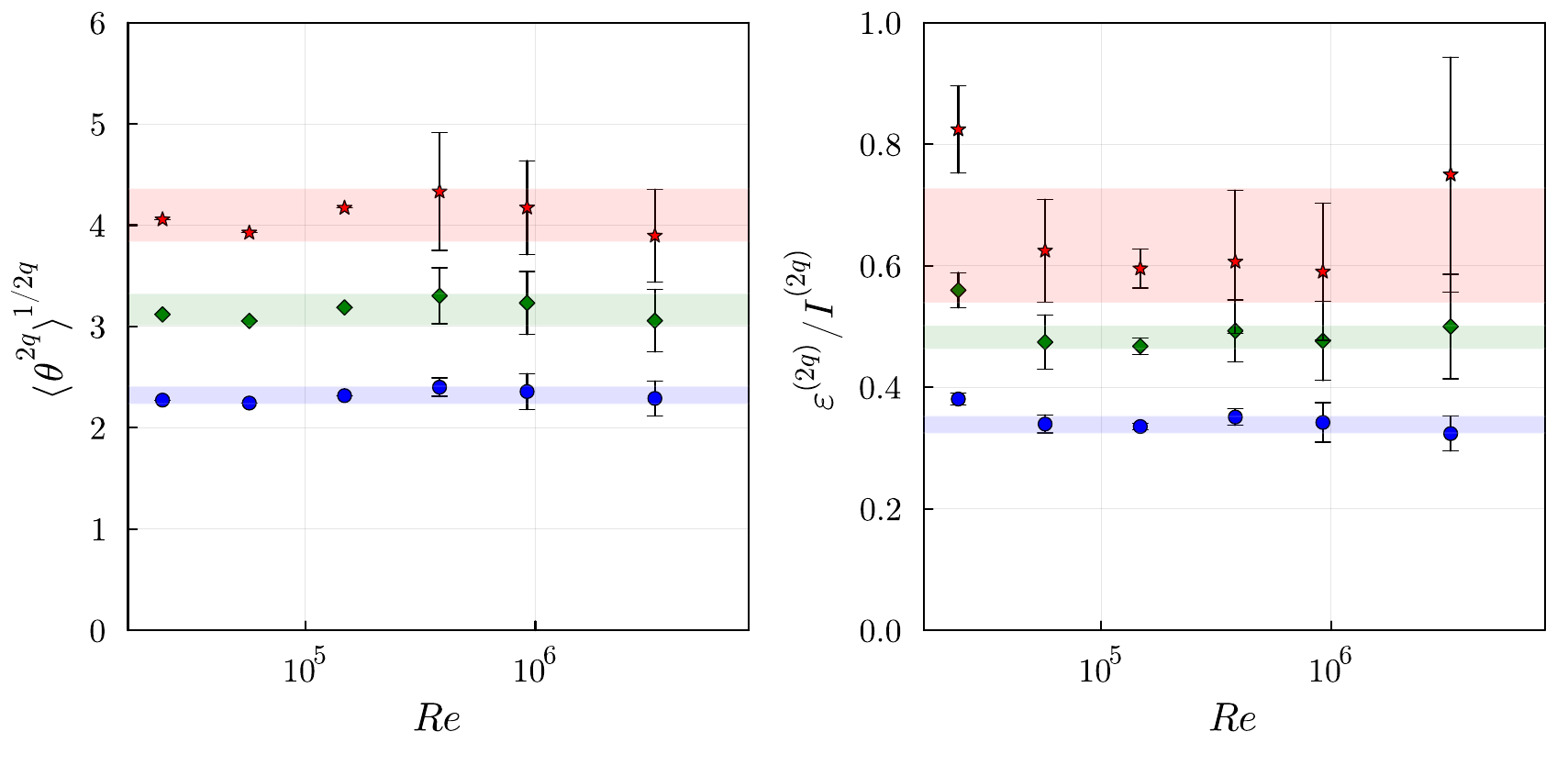}
\caption{\emph{Left:} Reynolds dependence of higher-order scalar moments. \emph{Right:} Corresponding viscous dissipations rates normalised by $I^{(2q)}=q^2\av{\theta^{2q-2}} \mI$ ---see Eq.~\eqref{eq:eps_2q}.}
\label{fig:4}
\end{figure}
Our simulations achieve a sufficiently strong form of statistical stationarity, where besides the scalar variance, the higher-order moments of $\theta$ also stabilise upon time-averaging and asymptote to finite values as $\Rey\to\infty$. This behaviour is illustrated in figure~\ref{fig:4}. 
These moments are associated to non-quadratic inviscid conservation laws (Casimirs), and are dissipated by both large-scale damping and viscosity. The steady-state  budget for $\av{\theta^{2q}}$ is
\begin{equation}
	\label{eq:eps_2q}
	\varepsilon^{(2q)}:={\nu\,\av{|\bnabla(\theta^{q})|^2}} ={q^2\,\av{\theta^{2q-2}}\mI} -\dfrac{\alpha q^2}{2q-1}{\,\av{\theta^{2q-1}\theta_\mf}}.
\end{equation}
Similarly to the quadratic case, the dissipation rates exhibit anomalous behaviours: one observes $\varepsilon^{(4)}\simeq 1.9 \mI\av{\theta^2}$, $\varepsilon^{(6)} \simeq 4.8 \mI\av{\theta^4}$, etc.

\subsubsection{K41 phenomenology in SQG turbulence}
\label{ssec:K41}
\begin{figure}
\includegraphics[width=0.99\textwidth]{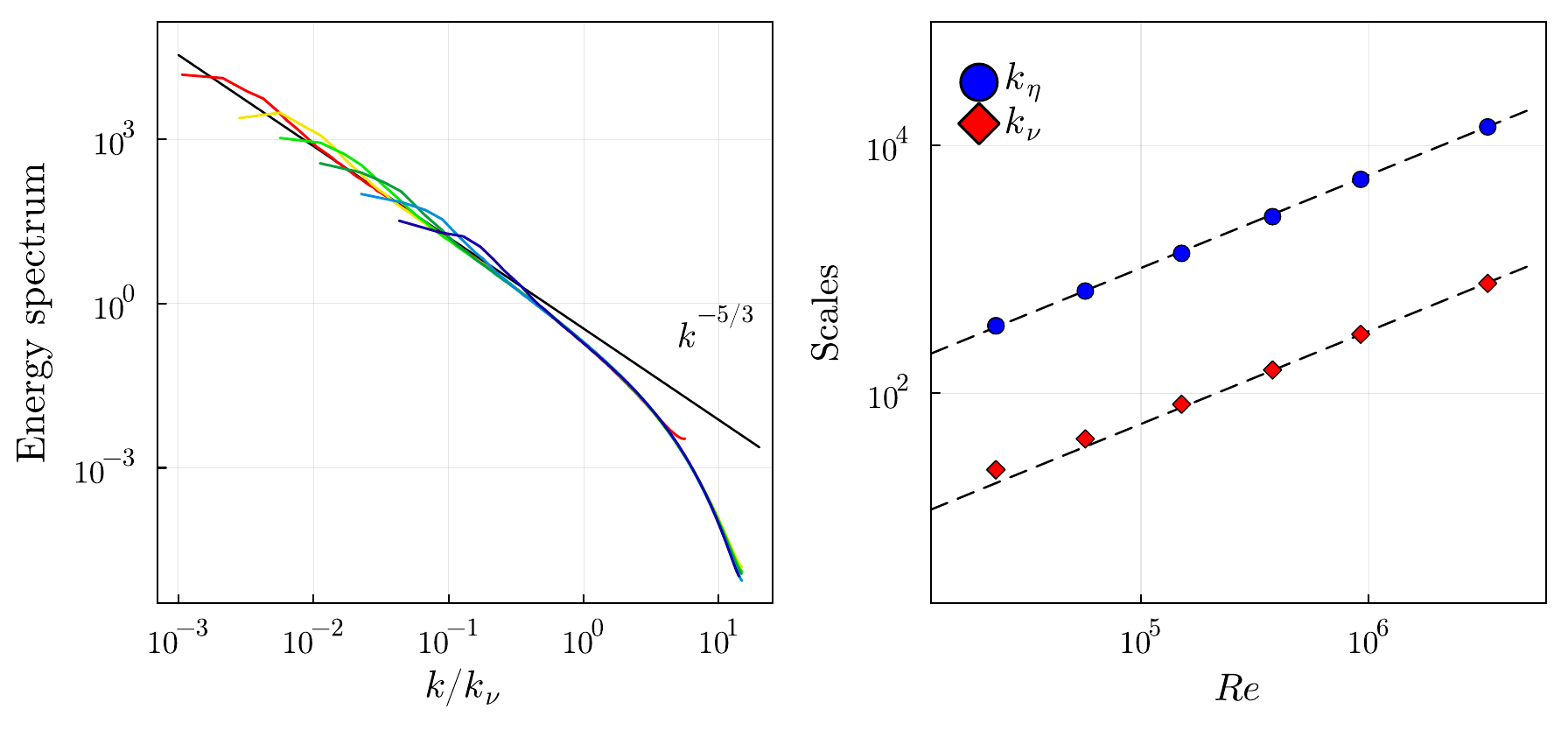}
\caption{\emph{Left}:~Energy spectra  observed from the six runs; the black dashed line indicates the Kolmogorov prediction $C_\mathrm{K}\varepsilon^{2/3} k^{-5/3}$, with a Kolmogorov constant $C_\mathrm{K}\simeq 0.35$. \emph{Right:}~ Reynolds number dependence of the viscous wavenumber $k_\nu$ and Kolmogorov wavenumber $k_\eta$.   See the text for definitions.}
\label{fig:5}
\end{figure}
Figure~\ref{fig:5} shows the power spectra from the six different runs.
Kolmogorov 1941 (K41) similarity theory predicts the development of a scaling regime 
$E(k) \approx C_\mathrm{K}\,\varepsilon^{2/3}k^{-5/3}$
 for $k_\mI \ll k \ll k_\nu$, with an increasing range as $\Rey$ grows. The observed spectral slope is slightly steeper than $-5/3$, suggesting the presence of intermittency corrections, as discussed in \S\ref{section1}. As seen in the right panel of figure~\ref{fig:5}, the viscous and Kolmogorov wavenumbers, $k_\nu$ and $k_\eta$, share a similar scaling $\propto\Rey^{3/4}$ with $k_\eta/k_\nu \simeq 10$. This supports the interpretation of $k_\nu$ as the dominant dissipation scale.
To summarise, the SQG K41 picture is mediated by the identifications,  holding up to possibly non-universal constant factors:
\be
	\varepsilon \propto \mI,\quad  k_\nu \propto k_\eta, \quad \av{\theta_\mf^2} \propto \ell_\mI^{2/3}\mI^{2/3}.
\ee
These identifications lead to $\gamma \simeq \varepsilon/k_\nu \propto \mI k_\eta \propto \Rey^{-3/4}$, consistent with the algebraic behaviour reported in figure~\ref{fig:3}. 


\section{Homogenous isotropic SQG turbulence (\hit)}
\label{section3}
From the spectral picture of K41 theory, one might think that the statistics of $\theta$ and $\bu$ are similar. This however dismisses the  presence of  physical-space structures, which are already apparent at the level of one-point statistics. 
A more explicit distinction between the two SQG fields emerges when considering two-point statistics, which explicitly relate  to cascade mechanisms. 
The SQG version of Yaglom's law ties  a mixed third-order structure function to the SQG dissipation, revealing non-trivial correlations between $\theta$ and $\bu$. This exact law provides a natural framework for evaluating finite-viscosity effects and identifying inertial-range statistics. 
After discussing the basic phenomenology of one-point statistics, we examine the validity of Yaglom's law in our numerical simulations and discuss its implications for the statistics of $\theta$ and $\bu$ ---focusing in particular on the various third-order structure functions.

\subsection{One-point statistics}
\begin{figure}
\begin{minipage}[c]{0.4\linewidth}
   \centering
   \raisebox{0\baselineskip}{\includegraphics[width=1\linewidth]{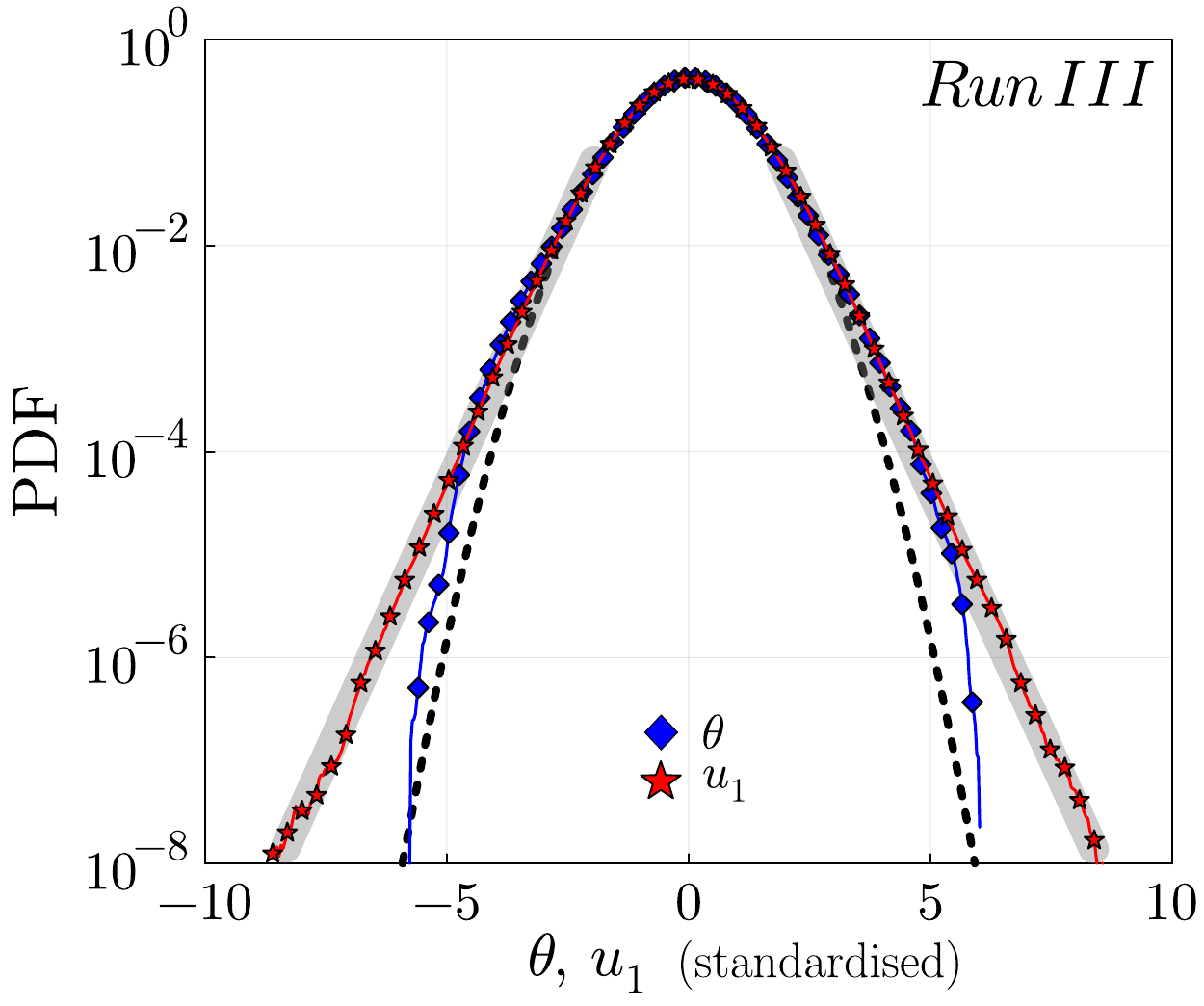}}      
  \end{minipage}
 \hfill
  \begin{minipage}[c]{0.59\linewidth}
   \centering
   \includegraphics[width=1\linewidth]{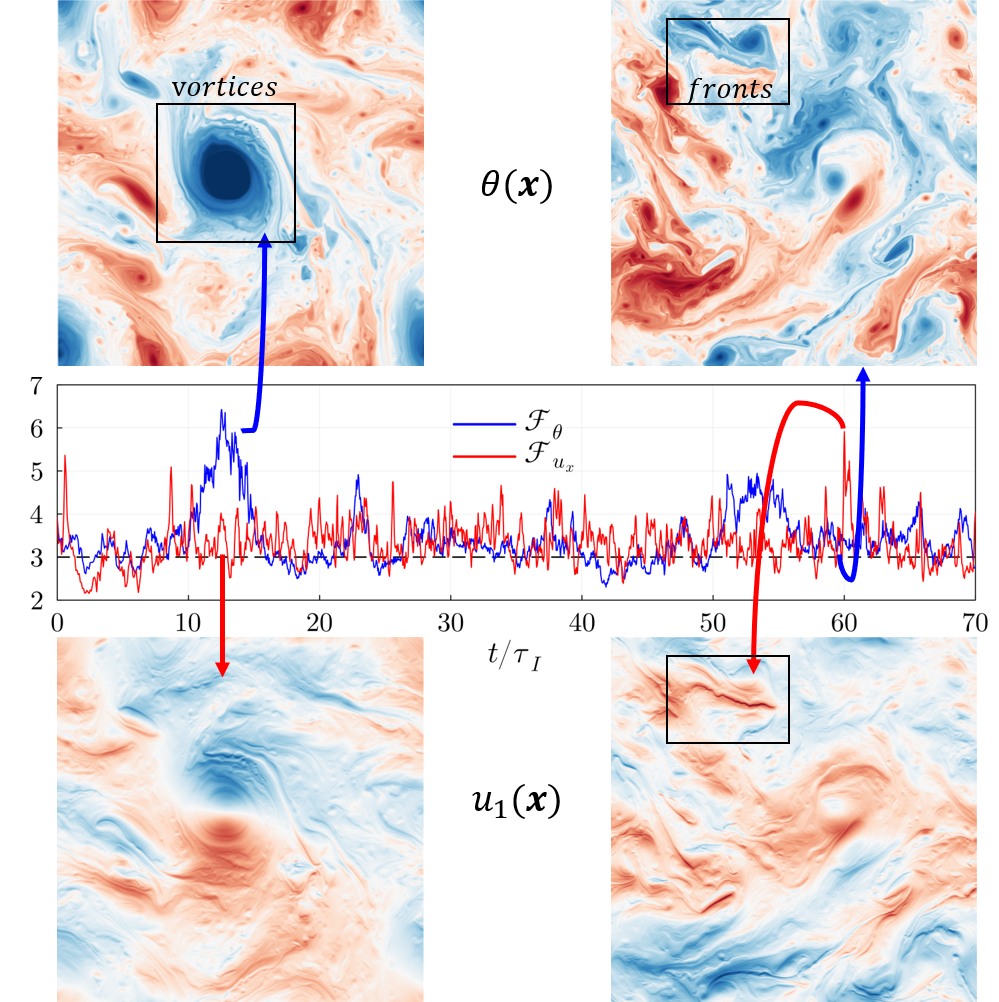}      
  \end{minipage}
\caption{\emph{Left:} Probability density functions (PDF) of the scalar field $\theta$ (\textcolor{blue}{$\mathbin{\blacklozenge}$}) and of the velocity component $u_1$ (\textcolor{red}{$\star$}) standardised to zero mean and unit variance. The black dashed line represents a Gaussian distribution, while the gray shaded areas show exponential tails $\propto e^{-2.4 u_1}$. \emph{Right:} Time series of the  spatially-averaged flatnesses  defined as
$\mathcal{F}_\theta(t) = \fint \theta^4 / (\fint \theta^2)^2$, $\mathcal{F}_{u_1}(t) = \fint u_1^4 / (\fint u_1^2)^2$, along with representative  fields configurations observed during two extreme fluctuations.}
\label{fig:1point}
\end{figure}
A basic phenomenology of SQG structures can already be grasped by comparing the one-point statistics of the scalar and velocity fields, as shown in figure~\ref{fig:1point} for Run III. 
The statistics in figure~\ref{fig:1point} are computed over a finite time-window $\simeq 300 \tau_\mI$:
Both  the scalar  and the velocity components exhibit heavy tails, with an exponential decay indicative of intense non-Gaussian events---more pronounced for the velocity field.
The  extreme events  in the two fields reflect different physical structures, which can be identified  through the time series of the spatially-averaged flatnesses, 
as shown in the right panel of figure~\ref{fig:1point}. 
At a qualitative level, large fluctuations in $\theta$  occur when  strong vortices form,
though these structures do not leave a clear footprint in the velocity fluctuations. 
By contrast, intense velocity fluctuations 
arise from the presence of long scalar fronts, along which the velocity becomes very large. In the inviscid case, such scalar filaments lead to  logarithmic divergences in the velocity field\,---\,see for instance \cite{juckes1995instability} and Appendix \ref{ssec:SQGprototypes}.
\subsection{Yaglom's law}
\label{ssec:Yaglom}
\begin{figure}
  \centering
  \includegraphics[width=1\linewidth]{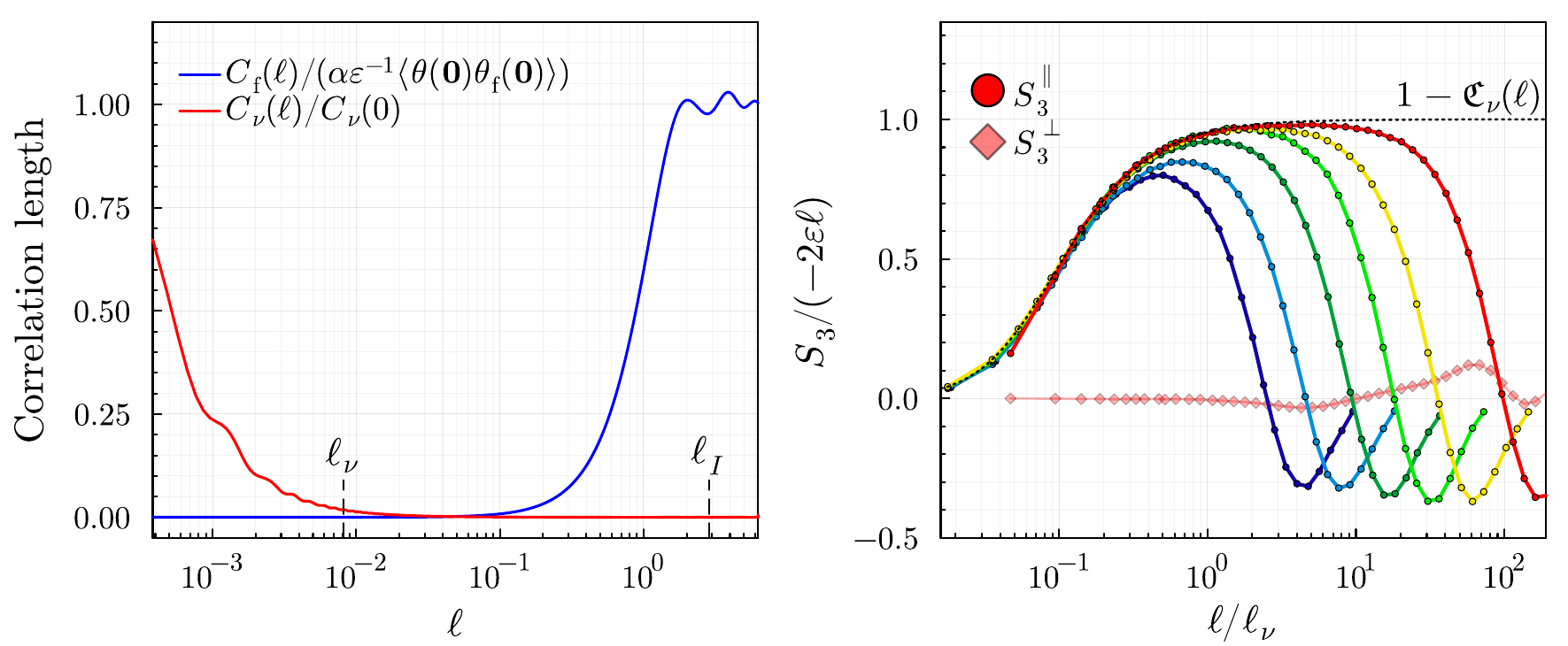}
  \caption{\emph{Left:} Correlations involved in the correction terms of Yaglom's law \eqref{eq:yaglom} obtained from Run VI. \emph{Right:}  Compensated structure function $S_3^\parallel = \langle\delta u_\parallel  (\delta\theta)^2\rangle/ (-2\varepsilon \ell)$,  as a function of the normalised separation $\ell/\ell_\nu$ for the various runs.  The black dotted line represents the viscous correction $1-\mathfrak C_\nu(\ell)$ in Eq.~\eqref{eq:yaglom}, measured for Run VI.  The diamonds indicate the perpendicular contribution to Yaglom's law.}
  \label{fig:7}
\end{figure}

To investigate two-point spatial statistics,  we define the increments of the velocity and scalar fields as 
$\delta \bu  =  \bu(\bx+\bl,t) -  \bu(\bx,t)$ and $\delta \theta  =  \theta(\bx+\bl,t) -  \theta(\bx,t)$. The longitudinal and transverse increments are denoted as $\delta u_\parallel =\delta \bu  \cdot \hat \bl$  and $\delta u_\perp = \delta \bu  \cdot \hat \bl^\perp$, with $\hat{\bl} = \bl/|\bl|$.
Assuming statistical stationarity, homogeneity, and isotropy, standard turbulence calculations yield an exact relation\,---\,Yaglom's law\,---\,for the mixed third-order structure function, $S_3(\bl) := \langle\delta \boldsymbol{u}\,(\delta \theta) ^2 \rangle$, which takes the form of the (vectorial) balance equation
\begin{eqnarray}
	\label{eq:yaglom}
	&&\boldsymbol{S}_3(\bl) = -2 \varepsilon \bl \left[1 -\mathfrak  C_\mf(\ell)-\mathfrak  C_\nu(\ell)\right],\quad \mathfrak C_{\mf,\nu}(\ell):= \dfrac{2}{\ell^2} \int_0^\ell \ell' C_{\mf,\nu}(\ell')\,\mathrm{d}\ell'\\
	&&\text{with } C_\mf(\ell):=\dfrac{\alpha}{\varepsilon} \left[ \av{\theta({\bf 0})\theta_\mf({\bf 0})}-\av{\theta({\bf 0})\theta_\mf(\bl)}\right] \text{ and } C_\nu(\ell):=\dfrac{\nu}{\varepsilon}  \av{\bnabla \theta({\bf 0})\bcdot \bnabla \theta (\bl)}, \nonumber
\end{eqnarray}
The right panel of figure~\ref{fig:7} shows the behaviour of these correlation terms.  
The inertial range is defined as the range of scales where the correction terms $\mathfrak  C_\mf$ and  $\mathfrak  C_\nu$ are negligible, that is $\lambda_{\nabla \theta} \ll \ell \ll \lambda_{\theta_\mf}$,  where the lower bound is set by the correlation length of the scalar gradient and the upper bound by that of the filtered field  $\theta_\mf$. In practice, figure~\ref{fig:7} reveals that these correlation lengths are of the order of $\ell_\nu$ and $\ell_\mI$, respectively.
 Within the inertial range,  equation~\eqref{eq:yaglom} prescribes the scaling of the third-order longitudinal and transverse mixed structure functions:
\be
 	S^\parallel_3(\ell) := \av{\delta u_\parallel\delta \theta^2} = -2\varepsilon \ell  \quad \text{and} \quad S^\perp_3(\ell)  := \av{\delta u_\perp\delta \theta^2} = 0.
\ee
The right panel of figure~\ref{fig:7} shows the longitudinal component of the structure function $S^\parallel_3$, compensated by $-2\varepsilon \ell$ for all six runs. The results reveal convergence towards an asymptotic profile, with a plateau in the range $\ell \gg \ell_\nu$ that expands as the Reynolds number $\Rey$ increases. At small scales $\ell \lesssim 0.5\ell_\nu$,  the  profile is dominated by viscous corrections, captured by $1-\mathfrak C_\nu(\ell)$.   
At  scales $\ell \gtrsim 20 \ell_\nu\simeq 0.1\ell_I$, the contribution from the large-scale damping can no longer be neglected, with deviations from Yaglom's scaling being at least of  the order of 10\%.
In summary,  our numerical results confirm the validity of Yaglom's law in the range $ 0.5\ell_\nu \lesssim \ell \lesssim 20  \ell_\nu$, providing a practical identification of the inertial range in SQG turbulence. For Runs V and VI, which have the highest resolution ($N^2=16,384^2$), this range  extends over slightly more than one decade. 

\subsection{Third-order structure functions}
While the mixed structure function $S_3$ prescribes a negative correlation $\propto \ell$ between the scalar and the velocity field, the behaviour of the third-order structure functions associated with each individual field are different.
The first distinction arises from symmetry considerations. In homogeneous isotropic SQG turbulence (SQG-HIT), both the scalar increment $\delta\theta$ and the transverse velocity increment $\delta u_\perp$  have symmetric distribution. This implies $S_3^\theta = \langle (\delta \theta)^3\rangle = 0$ and $S_3^{u_\perp} = \langle \delta u_\perp^3 \rangle = 0$ (see Appendix \ref{ssec:classical} for details).
\begin{figure}
  \centering
\includegraphics[width=0.99\textwidth]{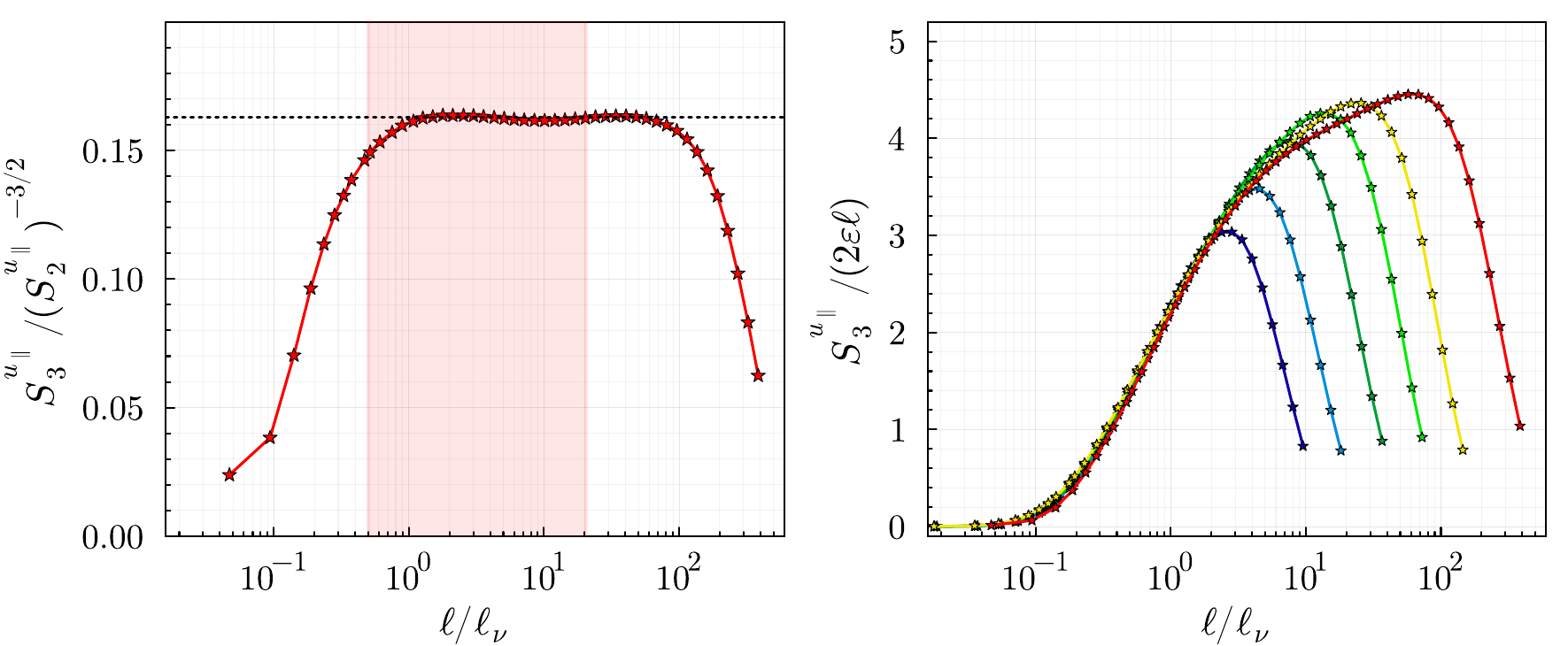}
  \caption{\emph{Left:} Skewness of the longitudinal velocity increments for Run VI. \emph{Right:} Third-order parallel structure functions for the various runs, compensated by Yaglom's scaling $2 \varepsilon \ell$.}
\label{fig:8}
\end{figure}
A more subtle observation concerns the statistics of longitudinal velocity increments, which exhibit positive skewness over the SQG inertial range (see the left panel of figure~\ref{fig:8}). Given that both $\bu$ and $\theta$ share the same direct cascade spectrum, this positive skewness is somewhat counterintuitive. 
Another intriguing feature is that Yaglom's scaling does not hold for the amplitude of $S_3^{u_\parallel}=\langle \delta u_\parallel^3 \rangle$. No clear plateau emerges in  the right panel of figure~\ref{fig:8},  where $S_3^{u_\parallel}$  is shown for Runs III to VI, normalised by Yaglom's scaling $2\varepsilon\ell$.
To quantify deviations from linear scaling, we use hereafter the local scaling exponents 
\be
\label{eq:local}
	\zeta_3^{u_\parallel}(\ell) = \dfrac{\mathrm{d} \log S_3^{u_\parallel}}{\mathrm{d} \log \ell}.
\ee
The left panel of
figure \ref{fig:9} reveals that in the SQG scaling range, all our runs exhibit a monotonous decay of $\zeta_3^{u_\parallel}$ 
from 1.8 to 1.1, indicating the absence of a well-defined power-law scaling range. For Run VI, we cannot entirely rule out a power-law behaviour with exponent $\zeta_3^{u_\parallel} \simeq 1.1$ over the range $5 \ell_\nu \lesssim \ell  \lesssim 50 \ell_\nu$,
but this barely overlaps with the previously defined SQG inertial range.
One interpretation is that the discrepancy comes from finite-viscosity effects, which  may influence $S_3^{u_\parallel}$ over a broader range than for $S_3^\parallel$.  Another interpretation is that  the apparent scaling range close to the forcing %
 scale  %
 is simply  a visual artefact. It would then reflect a large-scale inflection point due to the  finite damping coefficient that we used to maintain steady-state statistics.
While we lean towards the second interpretation, providing a clear-cut answer would require extending our study to tremendously large resolutions.

As a final comment, please observe that the skewness of the longitudinal velocity vanishes at small scales.  
This is a consequence of incompressibility and isotropy playing out in a  2D domain, leading to $\partial_1 u_1 = -\partial_2 u_2  \overset{\rm law}\sim  -\partial_1 u_1$ (see also Appendix~\ref{sym_cons}). 

\subsection{The kinematic nature of the skewness phenomenon}
\label{ssec:kinematicskewness}
\begin{figure}
  \centering
    \includegraphics[width=0.99 \linewidth]{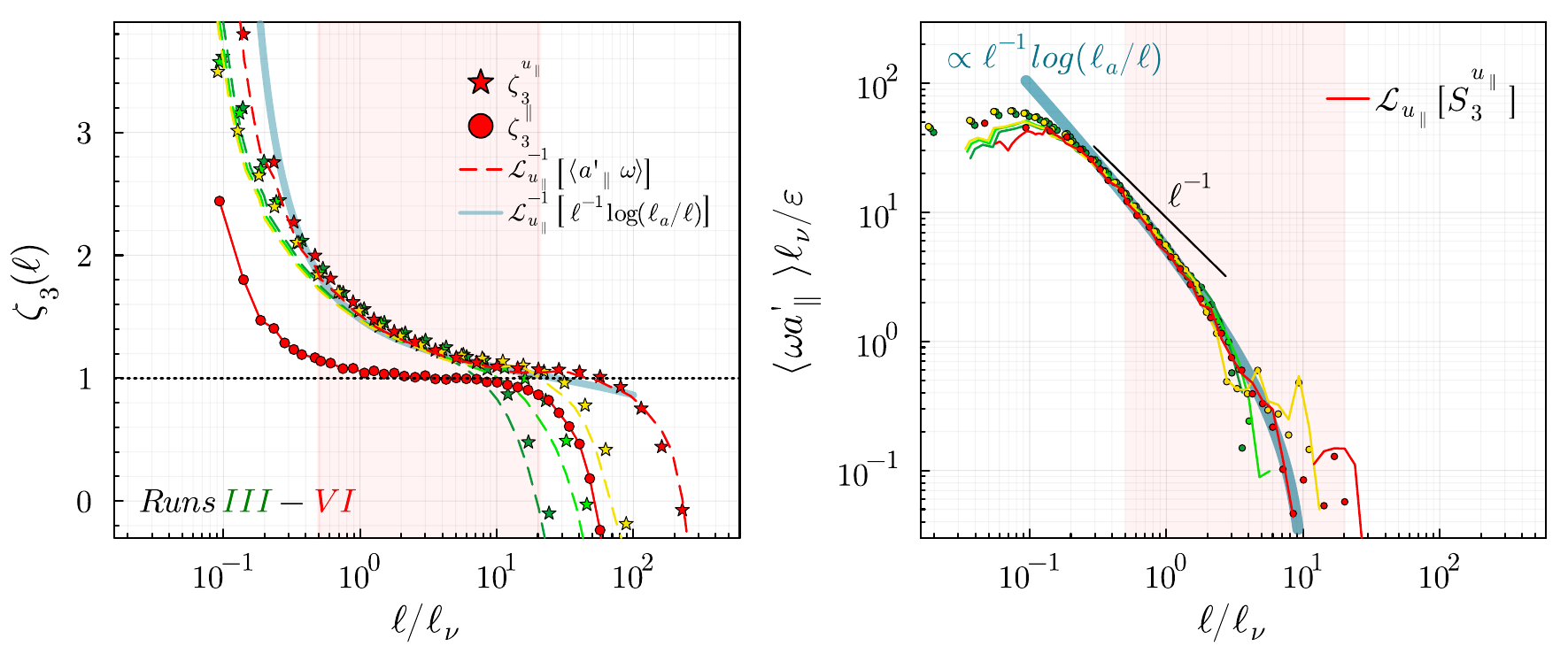}
  \caption{ \emph{Left:}   Scale-dependent scaling exponents for the third-order structure functions, determined through logarithmic derivatives.  \emph{Right:} Ageostrophic term. The solid lines represents Bernard's balance \eqref{eq:bernardbalance}.}
  \label{fig:9}
\end{figure} 
\label{fig:8}

At first glance, the positively skewed velocity increments observed in figure \ref{fig:8} contradict the direct cascade nature of the SQG transport and could be misleadingly interpreted as a velocity inverse cascade.
In  our setting of SQG turbulence, the energy flux is fully prescribed by the mixed structure function $S_3^\parallel$, which is negative and follows Yaglom's law~\eqref{eq:yaglom}. The velocity structure function $S_3^{u_\parallel}$ in turn does not enter by itself any simple scale-by-scale energy budget, and is necessarily compensated by an \emph{ageostrophic} contribution that counterbalances its sign \citep[see][]{capet2008surface}. This contribution arises from the integral identity  \eqref{NLOperators}  tying  $\theta$ to $\bu$, which provides a purely kinematic relation between the unmixed structure function ($S_3^{u_\parallel}$) and the mixed one ($S_3^\parallel$).

This relation
is mediated through a formula analogous to one pointed out by \cite{bernard1999three} for 2D Navier--Stokes equations. 
The formula \cb
is most conveniently written in complex notations, promoting in particular the vorticity $\omega=\bnabla^\perp \bcdot \bu$ and the scalar $\theta$ to functions $\Omega, \Theta$ defined on  the complex plane $\mathbb C$.
Writing $z = x_1 + i x_2 $, $U = u_1 + iu_2$, and $\delta U = U(z+\xi)-U(z)$ the (complex) increment over the separation $\xi \in \mathbb C$, the SQG version of Bernard's formula reads 
\begin{equation} \label{complexBernard}
   \dfrac{1}{3}\partial_{\xi\xi}\langle (\delta U)^3\rangle =  \partial_{\xi\xi^*}\langle \delta U (\delta \Theta)^2 \rangle - \langle A(0) \Omega(\xi)\rangle.
\end{equation}
Bernard's formula  involves the complex field $A = a_1 + ia_1$, associated to the ageostrophic vector  field $\boldsymbol{a} := (-\Delta)^{1/2}  [\bu \theta]- \bu \omega $.  
It turns out that the field ${\boldsymbol a}$ has a physical interpretation (see \S\ref{Lapeyre}).
In the complex setting,  Yaglom's law \eqref{eq:yaglom} entails the inertial range relation 
$\av{\delta U (\delta \Theta)^2} = -2  \varepsilon\xi$, implying that Eq. \eqref{complexBernard} there reduces to    $\partial_{\xi\xi}\langle (\delta U)^3\rangle =   - 3\langle A(0) \Omega(\xi)\rangle$.  In words, the third-order velocity statistics are blind to the direct cascade, in the sense that they  do not relate to anomalous dissipation but are only prescribed by the subleading terms featured in \eqref{eq:yaglom}.

The argument can be  freed from the complex setting, upon tediously observing that  Eq. \eqref{complexBernard} implies a balance between longitunal structure functions,  which takes the explicit differential form 
\be
	\label{eq:bernardbalance}
	\begin{split}
	& \mathcal{L}_{u_\parallel} S_3^{u_\parallel} = \mathcal{L}_{\parallel} S_3^\parallel + \av{a_\parallel(\ell \boldsymbol{e}_1)\,\omega(\boldsymbol{0})},\\
	\text{with } 
\mathcal{L}_{u_\parallel} f := -\dfrac{1}{12}\ell^{-2}&\partial_{\ell}  \left[\ell^{-1} \partial_{\ell}\left[\ell^5\partial_{\ell}\left( \ell^{-1} f \right)\right]\right], \quad \mathcal{L}_{\parallel} f= \dfrac{1}{4}\partial_\ell \left[\ell^{-1}\partial_\ell \left( \ell f\right)\right],
\end{split}
\ee
naturally involving the vorticity $\omega$ and the ageostrophic vector field $\boldsymbol{a}$.
Again, we observe $\mathcal{L}_{\parallel} [-2\varepsilon \ell] = 0$, so that the operator $\mathcal{L}_{\parallel}$ annihilates the leading-order term  appearing  in Yaglom's law.
In the inertial range, the scale-by-scale balance \eqref{eq:bernardbalance} therefore relies on the ageostrophic contribution in the right-hand side of equation~\eqref{eq:bernardbalance}  and the subleading term featured in Yaglom's law. 
This is confirmed numerically in the right panel of figure~\ref{fig:9},
showing  $\langle a_\parallel(\ell \boldsymbol{e}_1) \omega\rangle$  and $\mathcal{L}_{u_\parallel}S_3^{u_\parallel}$ matching  one another within the inertial range.
Besides, the ageostrophic contribution consistently departs from the dimensional scaling $\propto 1/\ell$, with an apparent log-corrected behaviour
\be
  	\langle a_\parallel(\ell \boldsymbol{e}_1) \omega\rangle \sim \dfrac{\varepsilon}{\ell}\log \dfrac{\ell_a}{\ell},\quad \ell <\ell_a \simeq 10 \ell_\nu
\ee
independent of the Reynolds number.  
This substantiates the idea that the absence of scaling in $S_3^{u_\parallel}$ arises from the contribution of $\mathcal{L}^{-1}_{u_\parallel} [\langle a_\parallel(\ell \boldsymbol{e}_1) \omega\rangle]$, as evidenced by the local slopes in the left panel of figure~\ref{fig:9}. 
Essentially, it shows that the velocity skewness in SQG originates from the kinematics imposed by the presence of the two fields $\bu$ and $\theta$, while the anomalous energy flux appears inconsequential.
We refer the reader to Appendix \ref{appendix:bernard} for the derivations of equations \eqref{complexBernard}	and \eqref{eq:bernardbalance}.

\color{black}
\subsection{Comments on the ageostrophic term  $\boldsymbol{a}$ }\label{Lapeyre}
\begin{figure}
  \centering
  \includegraphics[width=0.3 \linewidth,trim = 1.6cm 2.3cm 2.5cm 1cm, clip]{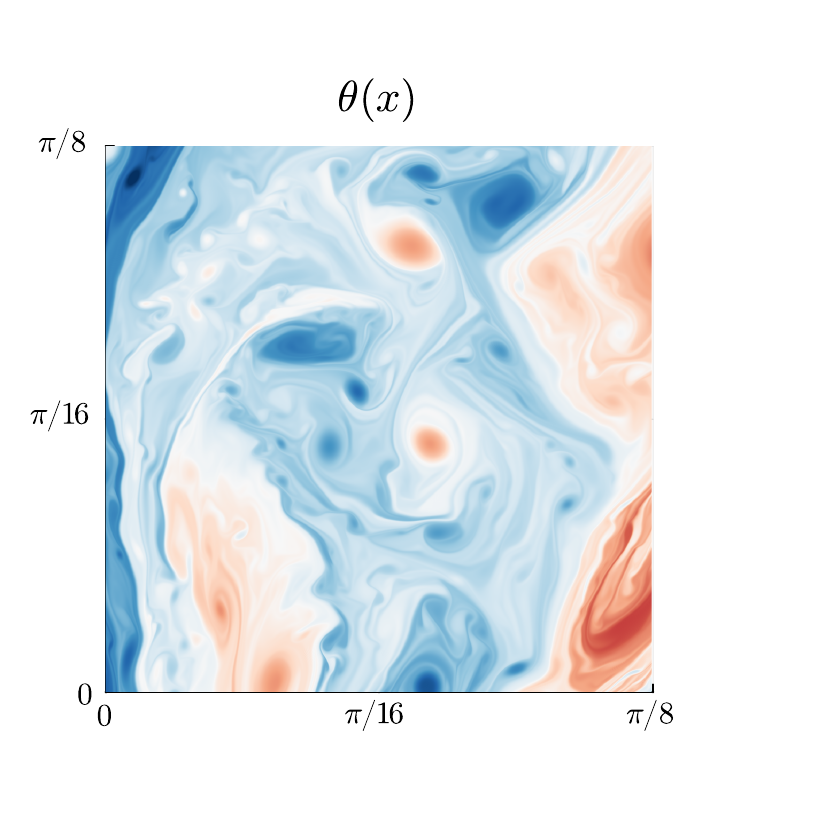}
  \includegraphics[width=0.3 \linewidth,trim = 1.6cm 2.3cm 2.5cm 1cm, clip]{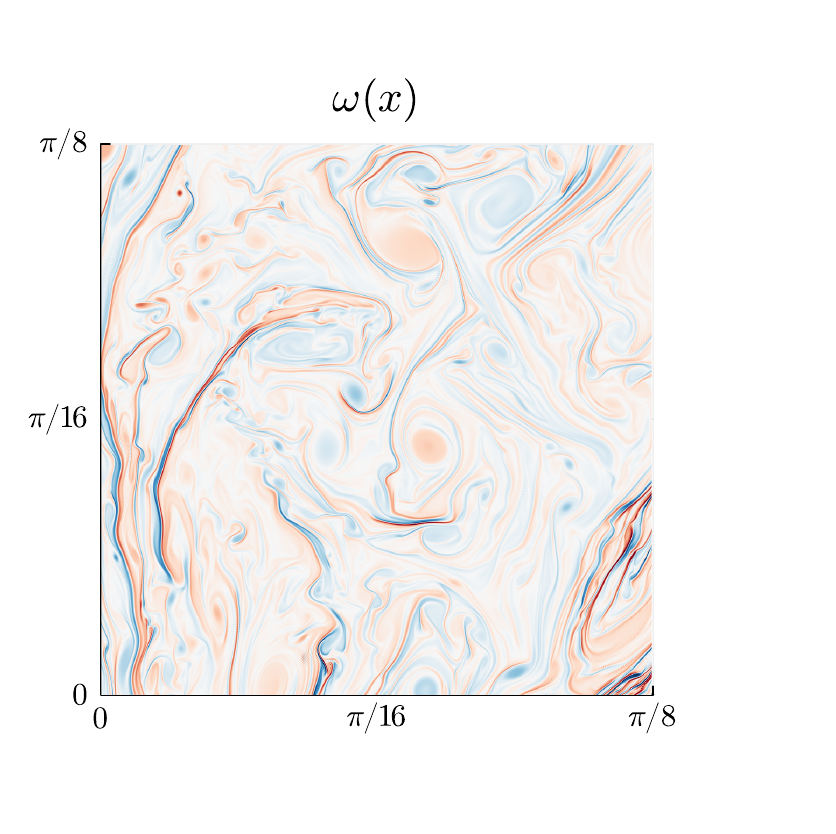}
  \includegraphics[width=0.3 \linewidth,trim = 1.6cm 2.3cm 2.5cm 1cm, clip]{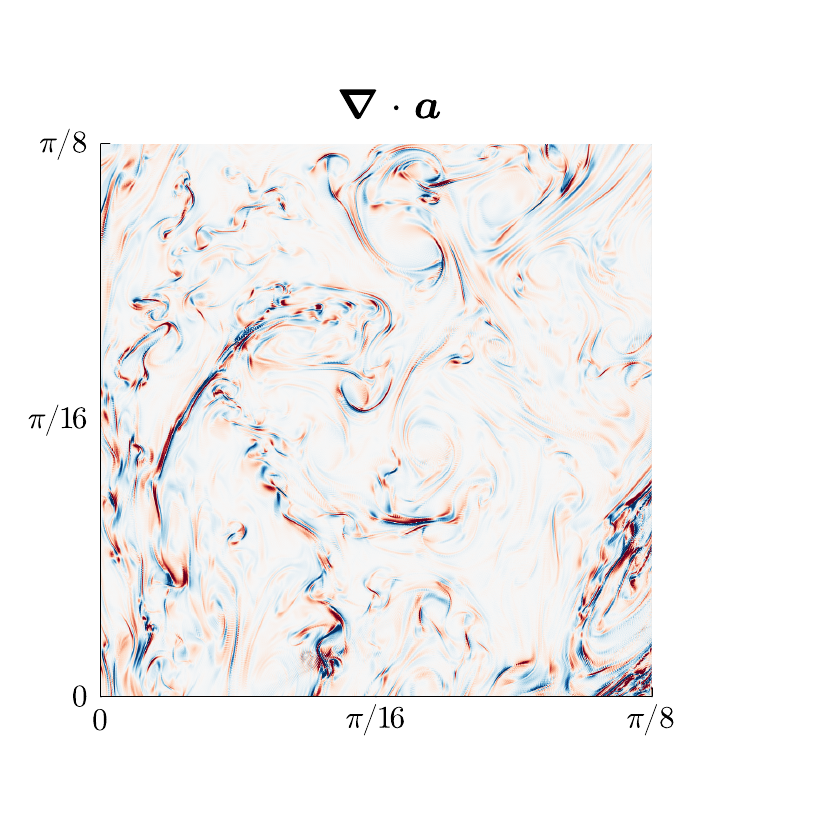}
  \includegraphics[width=0.065 \linewidth,trim = 2cm 2.3cm 10cm 2.3cm, clip]{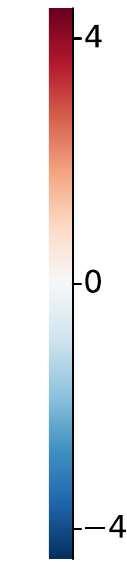}

  \caption{Snapshots of the scalar field $\theta$ (left), vorticity $\omega = \bnabla^\perp\bcdot\bu$ (center) and the divergence of the ageostrophic term $\bnabla\bcdot\ba$ (right), normalised to unit variance and zoomed over $(1/16)^2$ of the computational domain. (Data from Run V). }\label{fig:10}
\end{figure}

From a physical perspective, the ageostrophic field $\ba$ plays a central role in vorticity stretching mechanisms.
This becomes evident when examining the equation for vorticity $\omega = \bnabla^\perp \bcdot \bu$, derived from the SQG dynamics~\eqref{eq:sqg}:
\be
	\label{eq:sqg-w}
	\partial_t \omega + \bu \bcdot \bnabla \omega = -\bnabla  \bcdot  \ba+ \nu\Delta \omega + (-\Delta)^{1/2} F,\quad \ba:=  (-\Delta)^{1/2}  [\bu\,\theta]- \bu\,\omega 
\ee
Unlike in 2D Navier--Stokes turbulence, the enstrophy $(1/2)\fint \omega^2 $ is not conserved in SQG. Instead, its non-viscous production rate $\av{\omega \bnabla \bcdot \ba }$  is directly modulated by the ageostrophic velocity $\ba$.  
Figure~\ref{fig:10} illustrates the non-trivial correlations between $\bnabla \bcdot \ba$ and the SQG fields. In particular, $\bnabla \bcdot \ba$ exhibits strong values along the sharp fronts of a scalar filament, while remaining nearly zero elsewhere.

In geophysical terms, recall that the SQG framework governs the evolution of the surface buoyancy $\theta$ and serves as the boundary condition for three-dimensional potential vorticity inversion problem. Specifically, in the upper-ocean, the streamfunction $\psi$ satisfies a Poisson equation with mixed (Robin) boundary conditions, namely
\be
	\left(\partial_{11}+\partial_{22} +\partial_{33}\right) \psi = 0, \quad \left. \partial_3\psi\right|_{x_3=0}= \theta,\quad  \psi \underset{x_3\to -\infty}{\to} 0.
\ee
This system  arises as a  first-order perturbation of the primitive equations under the assumption of vanishing interior potential vorticity \citep{held1995surface}. Comparing equation~\eqref{eq:sqg-w} with formula~(3.a) in \cite{held1995surface}, we identify $\bnabla\bcdot \ba = -\left.\partial_3 u_3\right|_{x_3=0}$. In other words, up to a rotational component, $\ba$ represents the horizontal ageostrophic velocity field. Its regions of divergence (convergence) correspond to upward (downward) vertical motions.


\section{Intermittency}
\label{section4}
The distinct behaviours of the third-order structure functions indicate that the SQG fields $\theta$ and $\bu$, while non-trivially correlated, exhibit different scaling properties. They  should be treated as separate statistical entities. 
Besides, the apparent lack of systematic scaling raises the question of how to quantify intermittency in SQG turbulence. In this section, we report results on intermittency beyond third-order statistics, employing three different but consistent approaches to extract SQG anomalous scaling exponents, based on extended self-similarity, mixed structure functions and refined similarity. 

\subsection{Observations of intermittency}
\subsubsection{PDFs and flatness}
\begin{figure}
  \includegraphics[width=0.49\linewidth,trim = 0cm 0cm 0cm 0cm, clip]{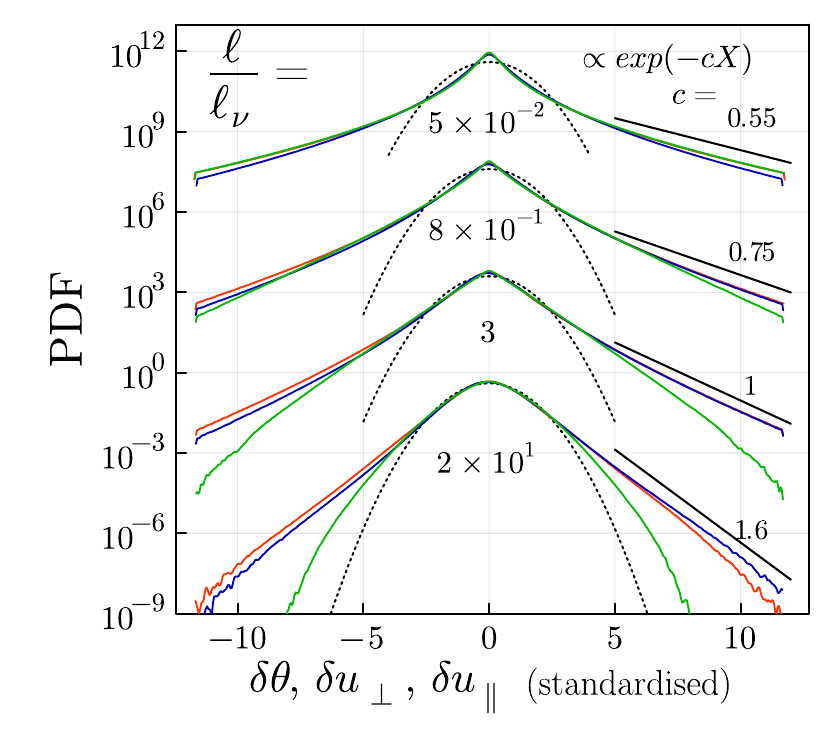}
   \includegraphics[width=0.49\linewidth]{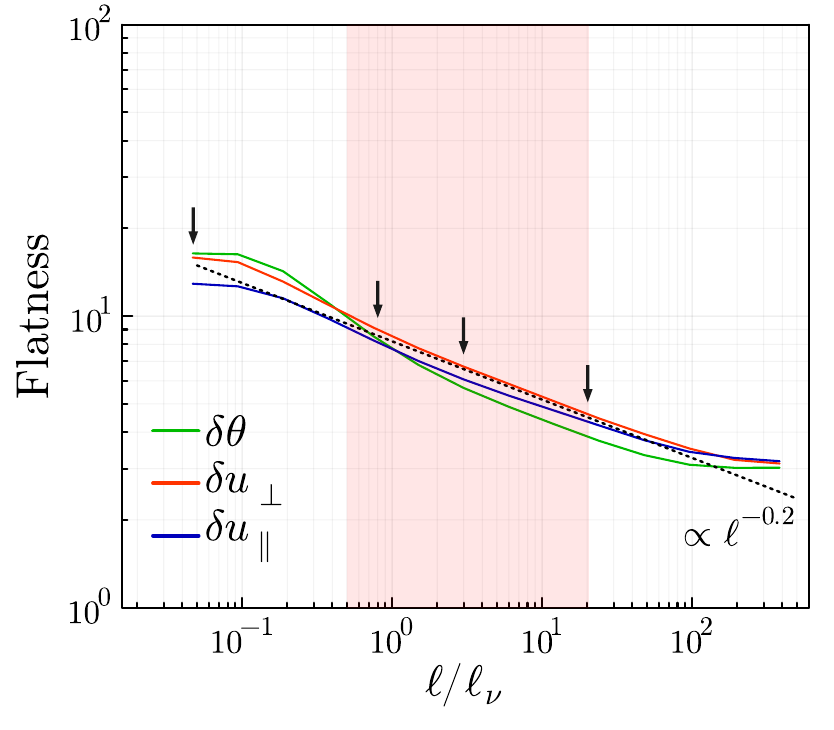}
  \caption{\emph{Left:} Probability density function of the scalar (green), longitudinal velocity (blue) and perpendicular velocity (red) increments for various separations, standardised to zero mean and unit variance, from Run~VI.   
Curves are shifted vertically for clarity. \emph{Right:} Corresponding flatnesses  as a function of the separation $\ell$. The vertical arrows indicates scales at which PDFs of the left panel were evaluated.} 
  \label{fig:11}
\end{figure}

The left panel of figure~\ref{fig:11} shows the probability density functions (PDFs) of standardised increments (zero mean, unit variance) of the scalar field, transverse velocity, and longitudinal velocity from Run VI. The distributions correspond to increment lengths spanning from the inertial range down to the grid resolution ($\Delta x=0.05 \ell_\nu$ for run VI). The scale-dependence  of the statistics  provides clear evidence of intermittency, and   highlights differences between the fields.
For large separations, the PDF of velocity increments  have exponential tails, with the longitudinal component $\delta u_\parallel$ displaying noticeable skewness.

\begin{figure}
  \centering
    \includegraphics[width=0.32\linewidth]{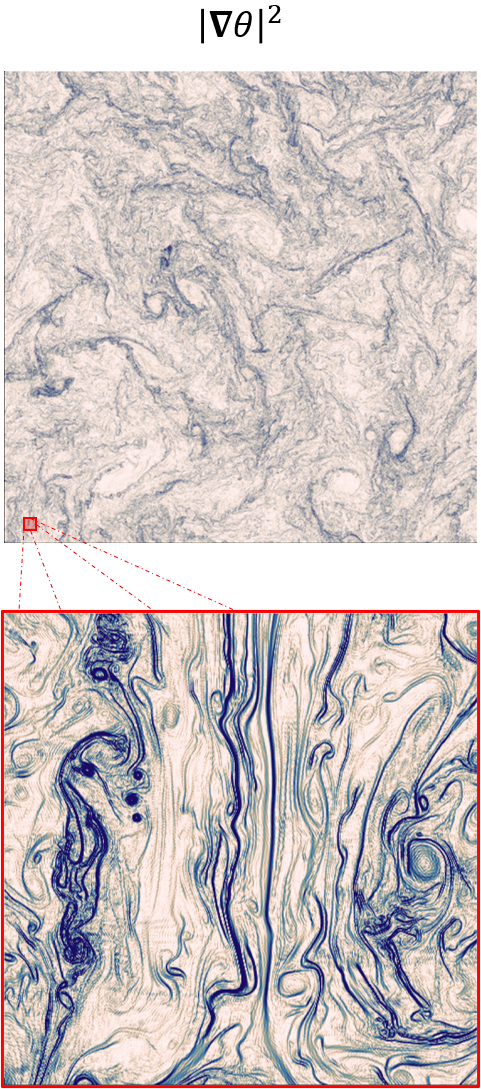}
    \includegraphics[width=0.339\linewidth,trim = -0.5cm 0cm 0cm 0cm, clip]{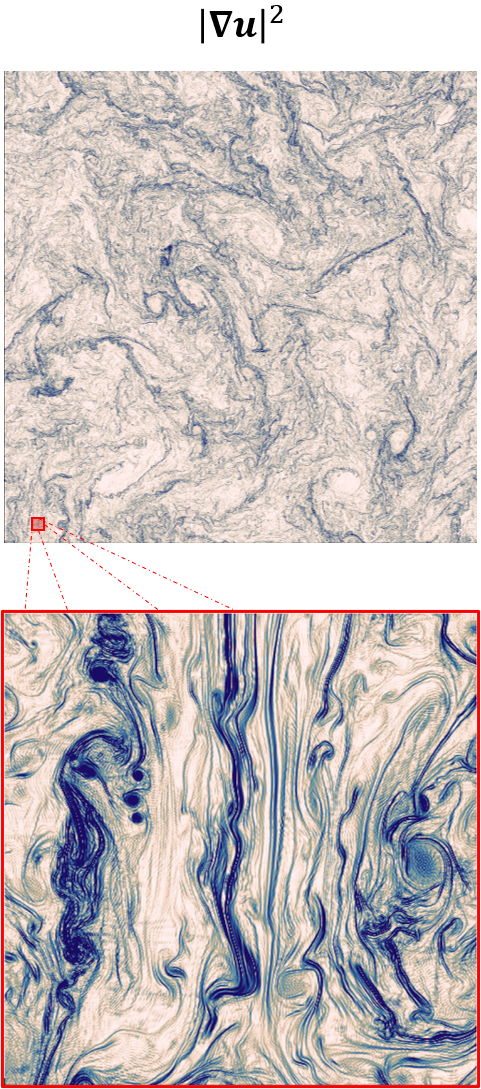}
    \includegraphics[width=0.124\linewidth,trim = -1cm 0cm 0cm 0cm, clip]{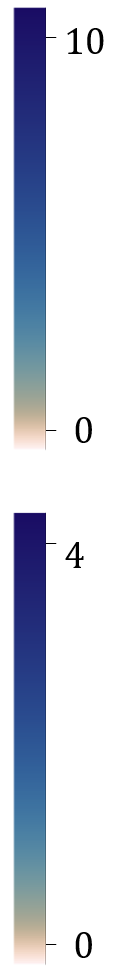}
    \caption{Typical snaphots of the squared  velocity gradient (left) and scalar gradient (right), normalised to unit variance, for Run VI.}
    \label{fig:dissipation_field}
  \end{figure}

As $\ell$ decreases,  the three PDFs flatten, with their tails becoming closer to stretched exponentials.
At the smallest scales $\ell \sim \Delta x$, the three distributions nearly collapse towards a symmetric profile, indicating that the gradient statistics of all three quantities are statistically alike (and unskewed).
This identification is captured visually in figure \ref{fig:dissipation_field}, where only a very trained eye could distinguish between the squared gradients of the velocity and scalar fields.
The widening of the PDF is quantified in particular by the flatnesses $F^\phi(\ell) = \av{\delta \phi^4}/\av{\delta \phi^2}^2$ ($\phi=\theta$, $u_\parallel$, and $u_\perp$),  which continuously increase  as $\ell \to 0$ as seen in  the right panel of figure~\ref{fig:11}. The inset displays the associated local slopes: While a scaling behaviour  
 $F^{u_\parallel},  F^{u_\parallel} \propto \ell^{-0.2}$ 
can be identified for the velocity increments   within the inertial range,  this is not the case for the scalar flatness $F^\theta$. These observations indicate that SQG intermittency extends beyond the third-order statistics discussed in \S~\ref{section3}.
  
\subsubsection{The lack of a (usual) scaling range}
\label{even}
\begin{figure}
\includegraphics[width=0.99\linewidth]{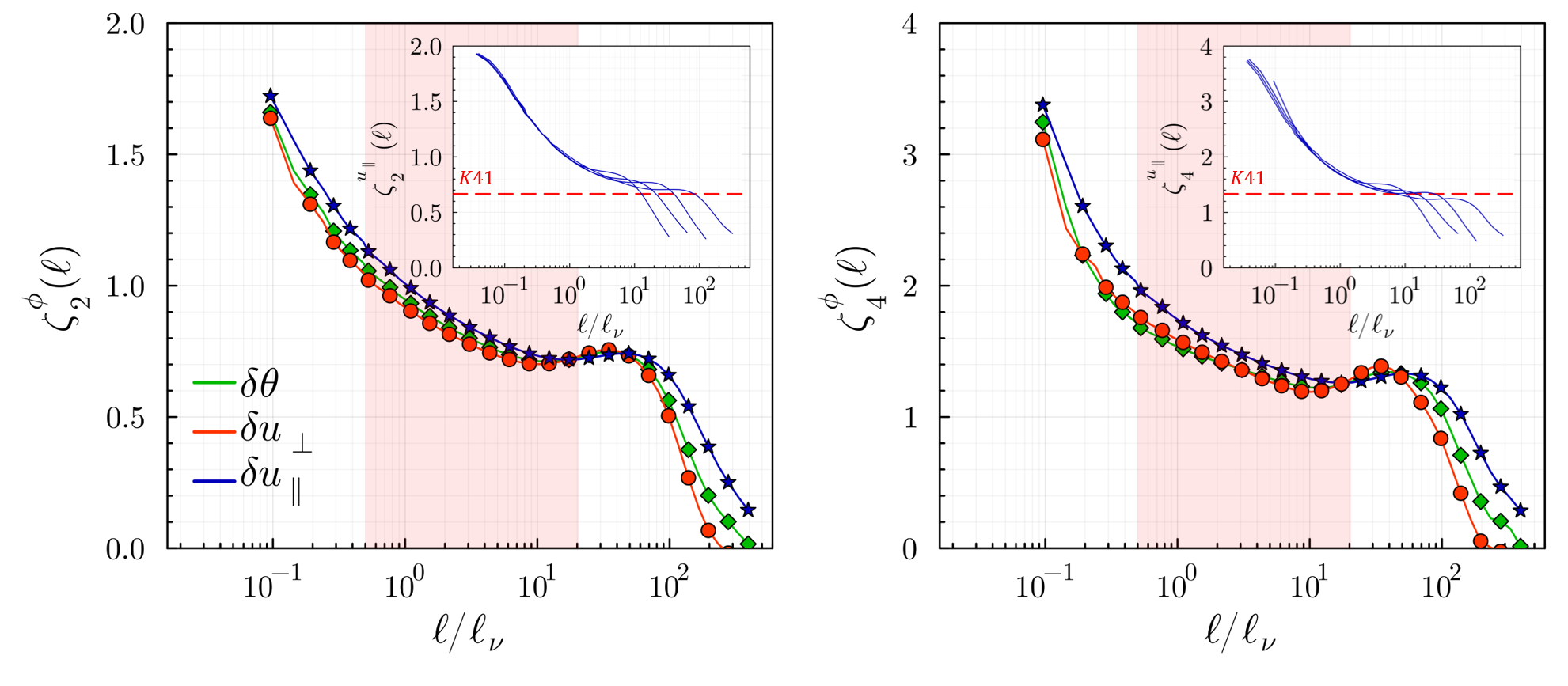}
  \caption{ \emph{Left:} Scale-dependent second-order exponents, $\zeta_2^\phi(\ell)$, for the SQG fields $\phi = \theta, \, u_\parallel, \, u_\perp$ from Run VI.  \emph{Inset:} Convergence of $\zeta_2^{u_\parallel}(\ell)$ with increasing Reynolds number $\Rey$. \emph{Right:} Same as the left  panel but for the fourth-order structure function. The dashed lines in the insets indicate the K41 values ($2/3$ and $4/3$).}
  \label{fig:12}
\end{figure}
In turbulence studies, scaling exponents are usually identified through the presence of an inertial range scaling for  the  structure functions of order $p$.
\be
	\label{eq:structure}
	S_p^{\phi}=\av{|\delta \phi|^p} \propto \ell^{\zeta_p^\phi},\quad \delta \phi := \phi(\bx+\ell {\bf e_1}) -\phi(\bx)
\ee
associated to a prescribed scalar field $\phi$.
In SQG turbulence,  the presence of well-defined scaling exponents is however \emph{not} a generic feature of single-field statistics. This feature is  worth emphasizing. Figure~\ref{fig:12} presents the behaviour with the separation $\ell$ of the local slopes  $\zeta_p^\phi(\ell) = \mathrm{d}\log S_p^\phi(\ell) / \mathrm{d}\log \ell$  for the second and fourth-order  structure functions  associated to the three SQG  fields ($\phi=\theta$, $u_\parallel$, and $u_\perp$). A  (usual) scaling range in the sense of equation \eqref{eq:structure} should translate 
into a plateau for the local exponent over the inertial range.

The main panels show the results for Run VI. At both orders, the local scaling exponents decrease monotonically with increasing scale within the inertial range---This trend is particularly pronounced for the longitudinal velocity increments.
The main observation is the absence of a well-defined plateau for $\ell \lesssim 20 \ell_\nu \simeq 0.1\ell_\mI$, suggesting that no inertial-range power-law scaling regime holds for any of the three SQG fields. 
The insets allow to assess the $\Rey$-dependency. They superpose the local scaling exponents $\zeta_2^{u_\parallel}(\ell)$ and $\zeta_4^{u_\parallel}(\ell)$ for  Runs III to VI. In both cases, the scaling exponents collapse at small scales, revealing a non-trivial scale-dependent master curve. 
The small plateau observed at intermediate scales can presumably be attributed to finite-size effects: It does not widen with increasing $\Rey$ and its value varies with the Reynolds number. Note that those apparent  asymptotic values do not coincide with the K41 similarity exponents, nor are bounded in any way by the latter.

The measurements shown in figure~\ref{fig:12} therefore provide strong evidence that neither the scalar field nor the velocity field exhibit true power-law scaling. In particular, this suggests that the energy spectrum slope is not a well-defined quantity: 
this could explain the difficulties  reported for instance by \cite{lapeyre2017surface} at consistently determining  scaling exponents in SQG turbulence. 
In spite of the lack of a scaling range in the usual sense, SQG intermittency can still be quantified through suitably defined (anomalous) scaling exponents. This is what we discuss next.

\subsection{Scaling exponents in SQG}
\label{scaling_sqg}
We now argue that SQG intermittency is structured by the transport properties of SQG turbulence, which in particular prescribe 
scaling laws for  mixed structure functions. 
Similarly to the third-order case, single-field statistics are not straightforwardly prescribed by these mixed structure functions. We however describe two ways to extract scaling exponents from single-field statistics:  extended self-similarity (ESS) and refined similarity (RSS). 
Both methods provide scaling exponents of order-$p$ consistently deviating from K41 similarity scaling $p/3$. The quadratic corrections are captured by Kolmogorov-Obukhov  log-normal dissipation model  $\zeta_p \simeq p/3 +\mu_{62}(p/3)(1-p/3)$ with $\mu_{62} = 0.16\pm 0.02$  up to order $p= 7$.

\subsubsection{Mixed structure functions}
\label{mixed_sf}
\begin{figure}
\includegraphics[width=0.49\linewidth]{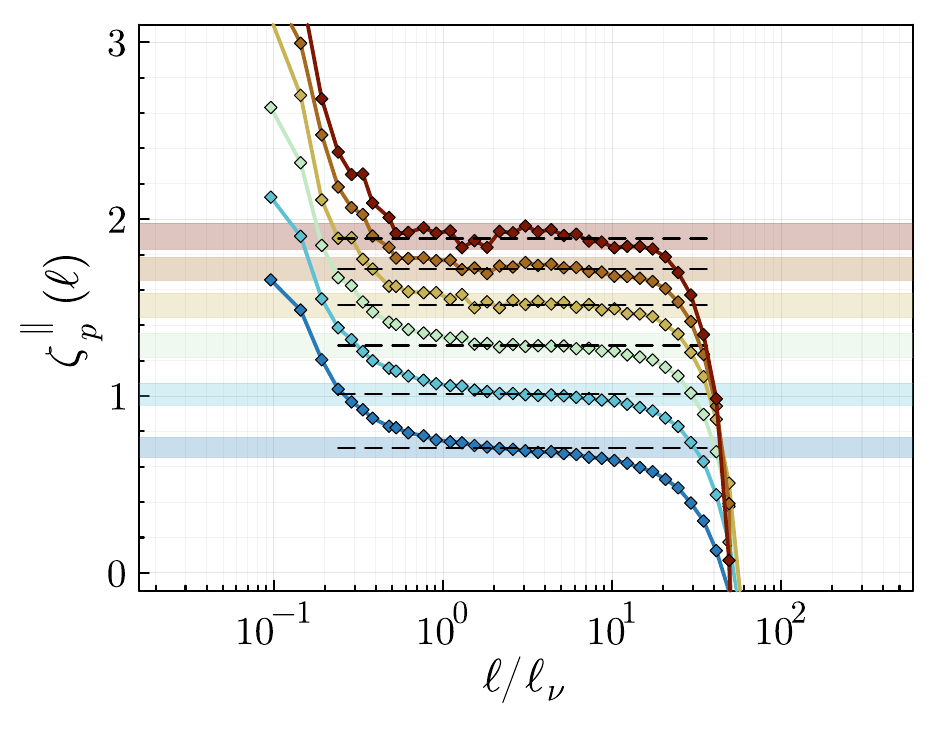}
\includegraphics[width=0.49\linewidth]{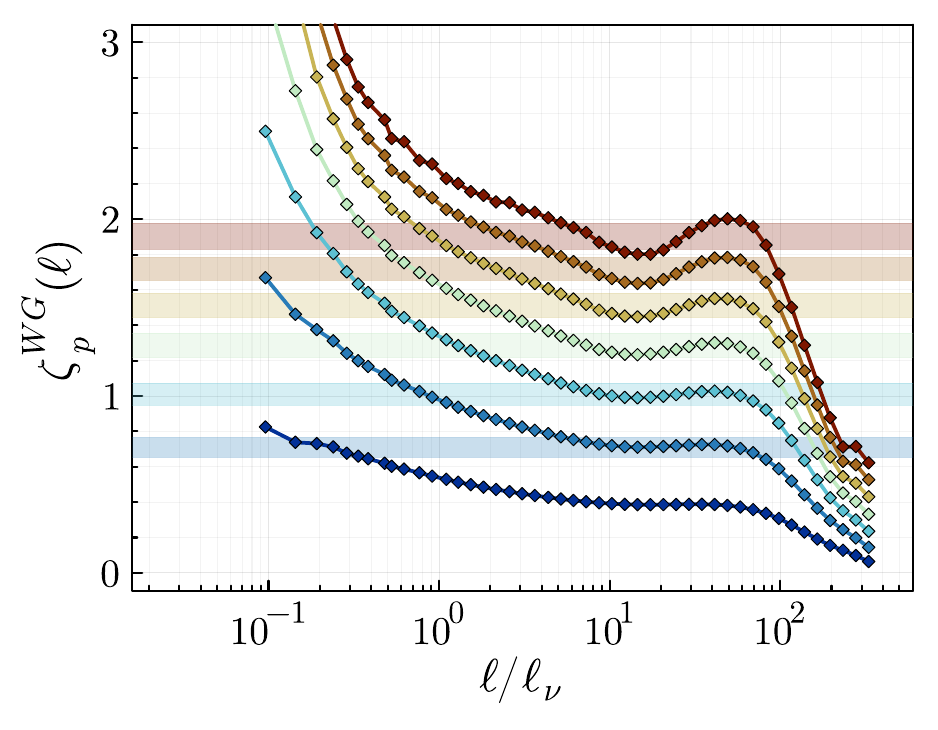}
   \caption{\emph{Left:} Local scaling exponents $\zeta_p^\parallel$ associated to the  mixed structure functions $\langle\delta u_{\parallel}|\delta \theta|^{p-1}\rangle$ for Run VI.  \emph{Right:}  Same as the left  panel but for the Watanabe-Gotoh mixed structure functions $\langle|\delta u_{\parallel}\delta \theta^{2}|^{p/3}\rangle$.  In both panels,  we show integer $p \in [2,7]$ (from blue/bottom to red/top). Dash lines indicate their inertial-range mean values and shaded areas their standard deviations. }
  \label{fig:Mixed_SF}
\end{figure}
The transport dynamics provides the most natural way to identify power-law scaling in the usual sense, reflecting the behaviour of mixed structure functions connected to conservation laws.  
The special case  $\langle \delta u_\parallel \,\delta \theta^2 \rangle \propto \ell$ associated to inertial-range constancy of the energy flux was extensively discussed   in \S\ref{section3}, in connection to Yaglom's law.
 The mixed structure functions associated to  inviscid conservation laws  of higher orders are defined by
\be 
	\label{eq:mixed-Sp}
	S_p^\parallel = \langle \delta u_\parallel |\delta \theta|^{p-1} \rangle. 
\ee
 For odd $p=2q+1>1$, the $S_p^\parallel$ are involved in the scale-by-scale budgets of the scalar moments $({1}/{2q}) \langle \theta^{2q} \rangle$, corresponding to conservation laws of inviscid SQG---Yaglom's law corresponds to $q=1$.
The left panel of figure~\ref{fig:Mixed_SF}  shows that the $S_p^\parallel$'s exhibit scaling in the usual sense, namely
\be	
	S_p^\parallel\propto \ell^{\zeta^{\parallel}_p}.
\ee
For $p \in [2,7]$, the local scaling exponents measured in Run VI  are nearly constant within an inertial scaling range increasing with $p$, hereby allowing for a faithful measurement of  $\zeta^{\parallel}_p$.

This type of scaling laws is similar to those observed in  passive scalar turbulence \citep{gauding2017high}.
In passive scalar turbulence, though, \cite{watanabe2004statistics} also identified scaling ranges for the  family of structure functions constructed as $S_p^{\rm WG}=\langle |\delta u_\parallel \delta \theta^2|^{p/3} \rangle$.   This observation does not extend to  SQG. The right panel of figure~\ref{fig:Mixed_SF} shows the local scaling exponent of the Watanabe-Gotoh  structure functions $S_p^{\rm WG}$, exhibiting monotonous decrease  within the inertial range. In particular, for $p=3$, the mixed structure function $\langle |\delta u_\parallel| \delta \theta^2 \rangle$ does not grow linearly with $\ell$, nor do any of its powers, and this contrasts with $S_3^\parallel = \langle \delta u_\parallel \delta \theta^2 \rangle \propto \ell$. In our interpreation, this discrepancy highlights the  role of velocity sign cancellations at producing scaling behaviour with the  (unsigned) longitudinal velocity increment $\delta u_\parallel$ driving the scale-by-scale transport.

\subsubsection{Extended self similarity}
\label{ESS}
\begin{figure}
  \includegraphics[width=0.49\linewidth]{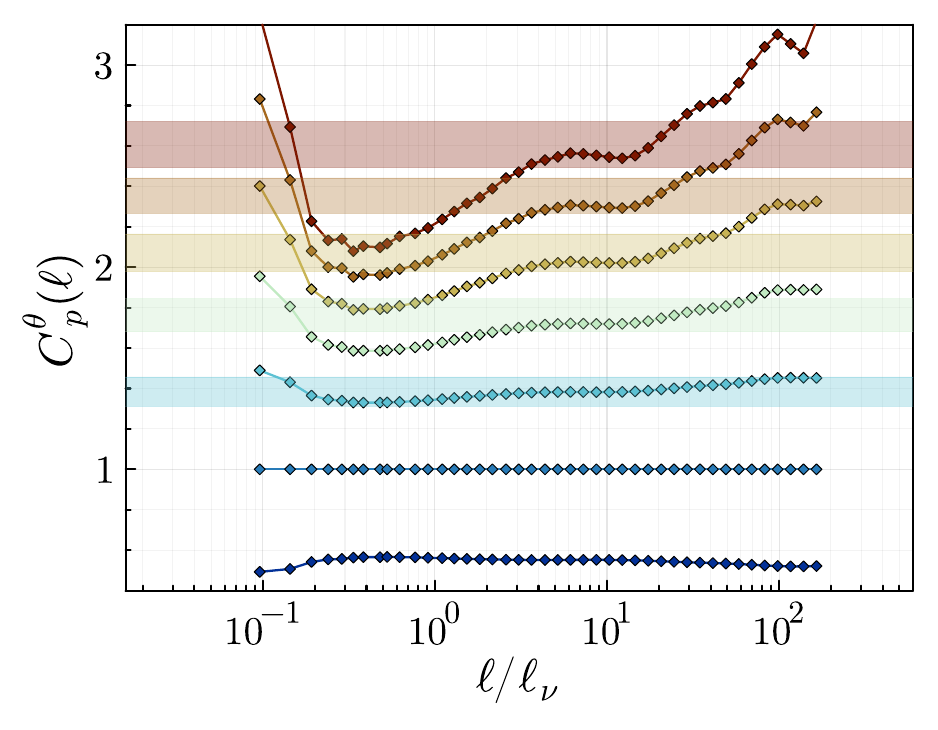}
  \includegraphics[width=0.49\linewidth]{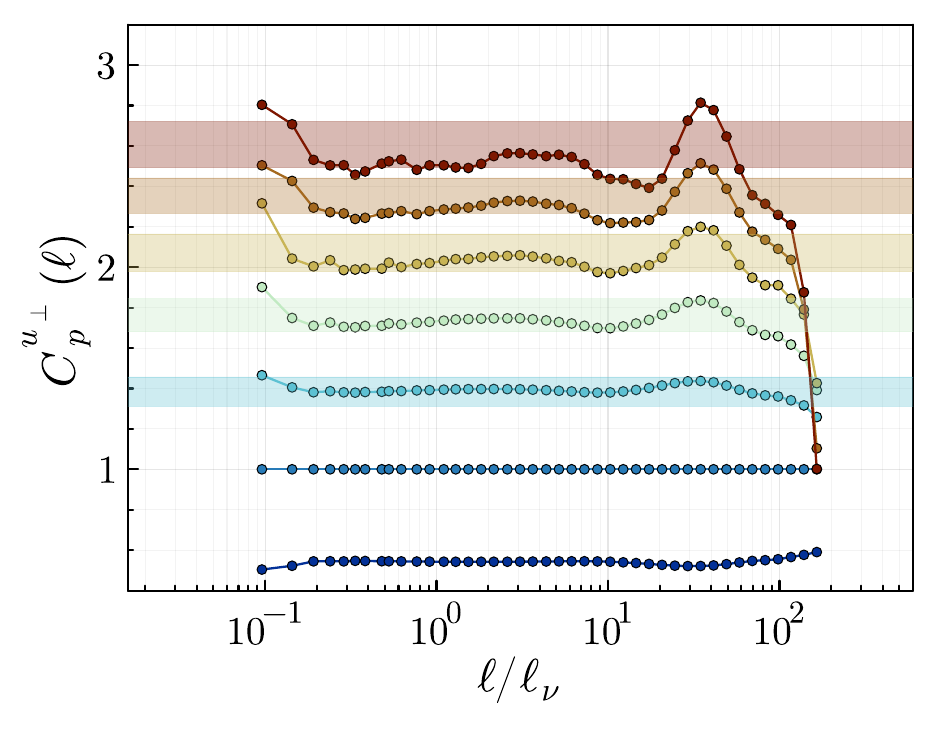}
  \includegraphics[width=0.49\linewidth]{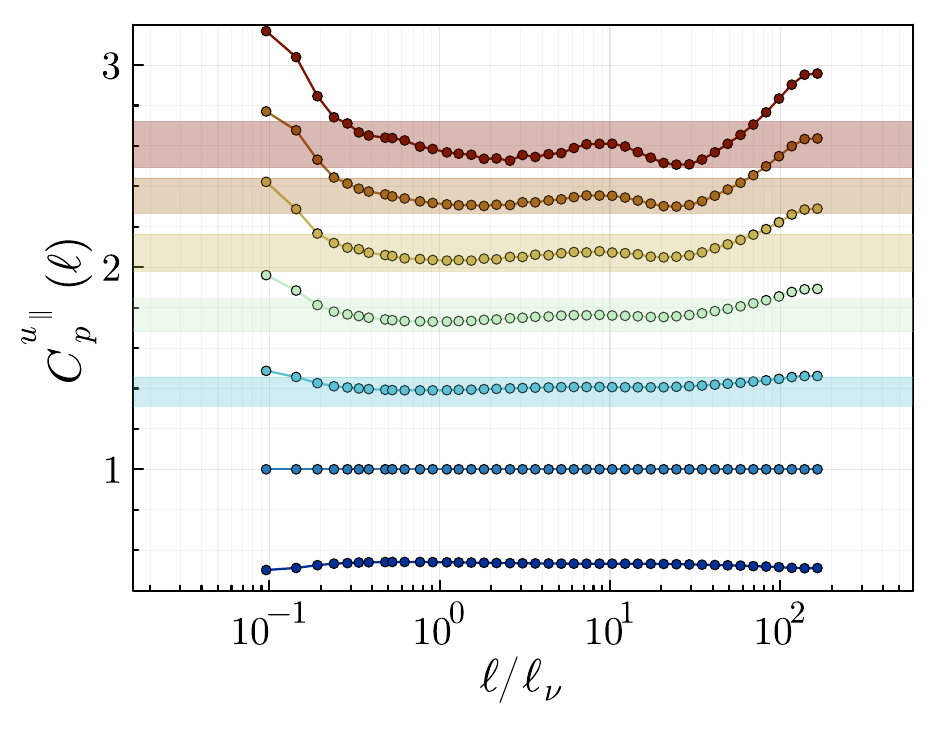}
  \includegraphics[width=0.49\linewidth]{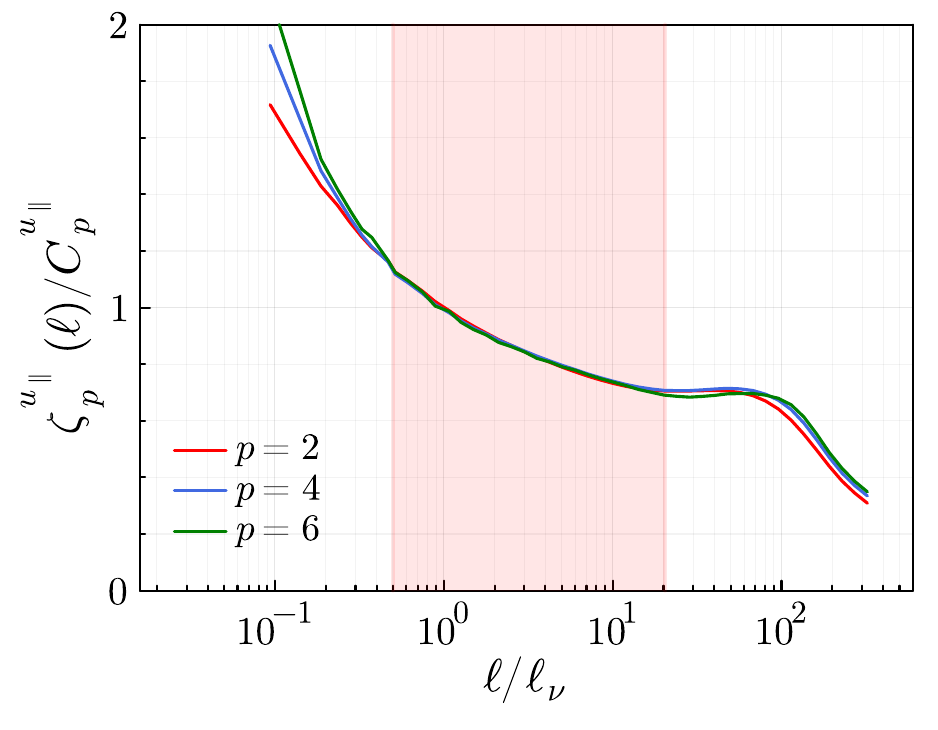}
 \caption{Extended self-similarity for integer orders $p \in [1,7]$ (from blue to red symbols) for the three SQG fields (left: $u_\parallel$, center: $\theta$, and right: $u_\perp$). The coloured areas represent the values obtained in \S\ref{mixed_sf} using mixed structure functions. The bottom right panel shows the scale-dependent exponent $\zeta_p^{u_\parallel}$ normalised by  $C_p$.}
  \label{fig:ESS}
\end{figure}
One way to extract power laws directly from single fields statistics is given by
Extended Self-Similarity (ESS) \citep{benzi2023lectures}. It consists in expressing structure functions in terms of a lower-order one in order to extend the range of scales over which power-law behaviour is observed.    
This strategy was introduced by \cite{benzi1993extended2,benzi1993extended} and proves remarkably efficient to extract scaling exponents in 3D homogeneous isotropic turbulence, even at relatively low Reynolds numbers. Instead of looking for pure power laws, ESS assumes a self-scaling property for the structure functions, namely
\be
	\label{eq:ESS}
		S_p^\phi \propto (S_2^\phi)^{C^\phi_{p}} \quad (\phi=\theta, u_\parallel, u_\perp),
\ee
with the exponents $C^\phi_p$ a priori different for each SQG field.
The presence of scaling ranges in the usual sense \eqref{eq:structure} implies ESS with 
$C_{p}^\phi = \zeta^\phi_p/\zeta^\phi_2$. The lack of usual scaling ranges does not however preclude the self-scaling property \eqref{eq:ESS}. For example, assuming a generalised scaling of the form $S_p^{\phi}(\ell) \propto \left[\ell f(\ell)\right]^{\zeta^\phi_p}$,  ESS would still hold, with the exponent $C_{p}^\phi = \zeta^\phi_p/\zeta^\phi_2$ characterising the dominant power law contribution \citep[see][]{benzi1993extended2}. 

The ESS analysis of our SQG runs is summarised in figure \ref{fig:ESS},  showing the scale-dependent 
ESS exponents $C_p^\phi(\ell)= \zeta^\phi_p(\ell)/\zeta^\phi_2(\ell)$  at integer orders $p \in [1,7]$, constructed from 
the scale-dependent exponents of the structure functions.
The presence of an ESS scaling range   translate into the $C_p^\phi(\ell)$ being constant over an inertial range of scales. 
While the scalar increments  exhibit no clear ESS scaling range, the velocity structure functions do. The local slopes of both longitudinal  and transverse  structure functions fluctuate around constant values $C_p^{u_\parallel}$ and $ C_p^{u_\perp}$ over a broad range of scales $0.1\ell_\nu \lesssim \ell \lesssim 40 \ell_\nu$, extending beyond the range over which Yaglom's law is observed. 
The bottom-right panel of figure \ref{fig:ESS} points out that when  normalised by their  ESS exponent $C^{u_\parallel}_p$,
 the scale-dependent exponents $\zeta^{u_\parallel}_p(\ell)$  of the velocity field collapse to a profile which is largely independent of the order: This is another way to substantiate the relevance of the ESS relation \eqref{eq:ESS}, showing that the quantities  $ ({S^\phi_p})^{1/C^\phi_p}$ are independent of $p$.

\subsubsection{Refined similarity}
\begin{figure}
\centering
  \begin{minipage}{0.49\textwidth}
   \includegraphics[width=\linewidth]{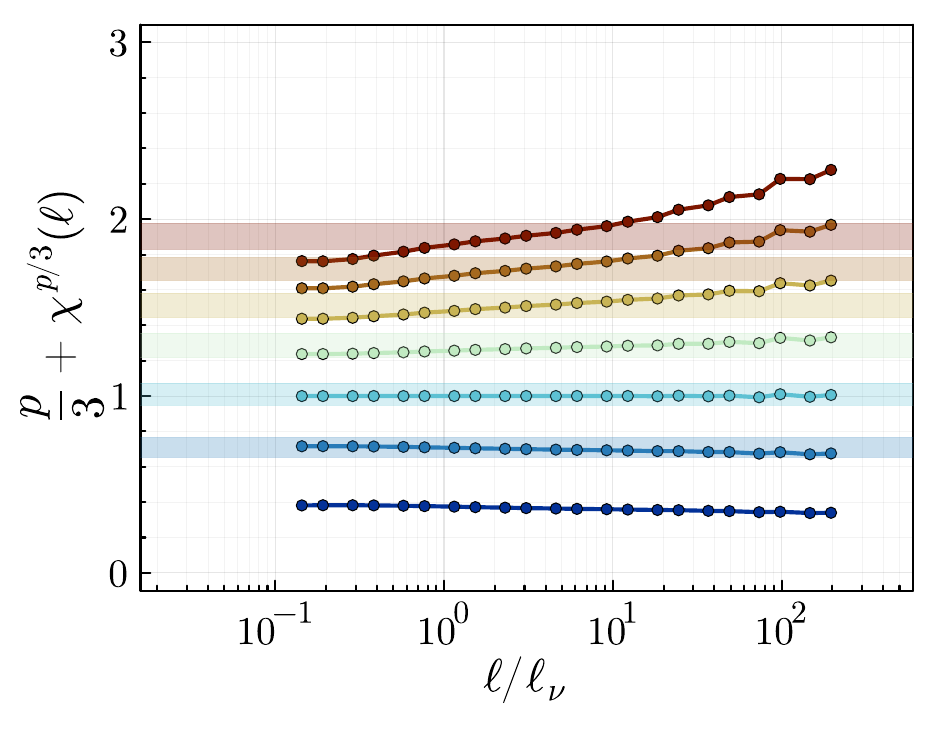}
   \end{minipage}
  \begin{minipage}{0.49\textwidth}
   \includegraphics[width=\linewidth]{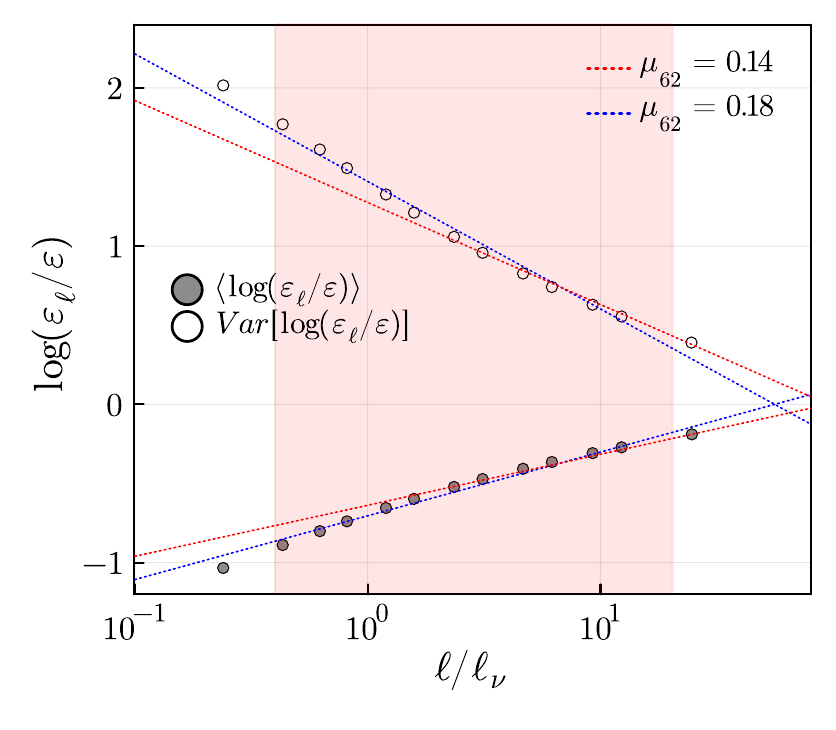}\\
   \includegraphics[width=\linewidth]{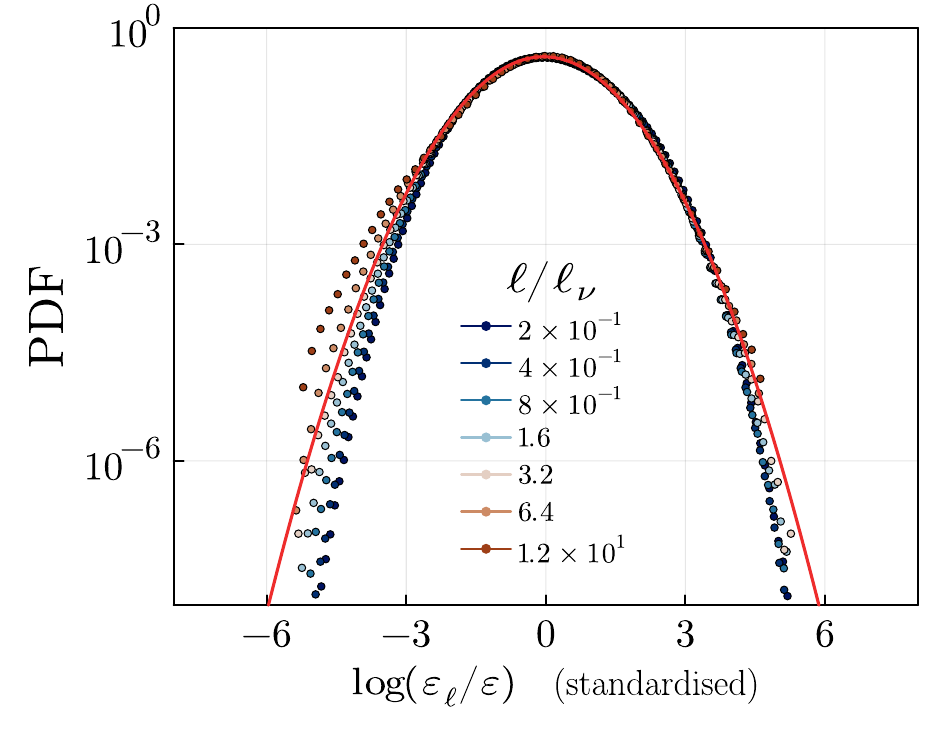}
   \end{minipage}
  \caption{\emph{Left:} Similarity exponents for integer values of $p \in [1,7]$. The shaded areas show the values for the mixed exponents.
  \emph{Top right:} PDF of the logarithm of the coarse-grained dissipation field standardised to unit variance and zero mean  \emph{Bottom right:} Mean and variance of $\log(\varepsilon_\ell/\varepsilon)$ as a function of the coarse-graining scale.
  }
  \label{fig:RSS}
\end{figure}

The refined similarity  theory of \cite{kolmogorov1962refinement} for 3D turbulence ties  deviations from self-similarity
to  heterogeneities in  the  scaling behaviours  for the coarse-grained dissipation field
\begin{equation}
\label{eq:epsilon_ell}
\varepsilon_\ell(\boldsymbol{x},t) :=  \frac{4}{\pi\ell^2} \int_{|\boldsymbol{r}|<\ell/2}\varepsilon (\boldsymbol{x} + \boldsymbol{r}) d^2\boldsymbol{r},\quad\text{with } \varepsilon(\bx,t) = \nu |\bnabla\theta|^2.
\end{equation}
Refined self-similarity suggests  the phenomenology $\av{\delta \phi^p}\sim \ell^{p/3}\langle\varepsilon_\ell^{p/3}\rangle$. 
 If the dissipation field is multifractal,  the moments of the coarse-grained field should behave as power laws 
$\langle\varepsilon_\ell^{q}\rangle \propto \ell^{\chi_q}$  \citep[see, \textit{e.g.},][]{frisch1995turbulence}, so that
\be
	\label{eq:bridge}
	 \av{\delta \phi^p} \propto \ell^{\zeta_p^\phi} \ \text{ with } \ \zeta_p^\phi = p/3+\chi_{p/3}.
\ee
In SQG,  though, such bridging relations do not hold---They would  imply scaling in the usual sense, which we recall is not observed (see \S\ref{even}). Still, our numerical simulations suggest that the dissipation field is multifractal and that the moments of dissipation behave as power laws:
This allows us  to use equation \eqref{eq:bridge} to \emph{define} the refined similarity scaling exponents as $\zeta_p^\phi= p/3+\chi_{p/3}$.
The left panel of figure~\ref{fig:RSS} shows the local similarity  exponents $p/3+\chi_{p/3}(\ell)$, with $\chi_{p/3}(\ell) = \mathrm{d} \log\langle\varepsilon_\ell^{p/3}\rangle/\mathrm{d}\log \ell$. 
The  exponents  remain approximately constant across all scales for $p\leq 5$, and vary by at most 10\% within the inertial range for $p=6$ and $7$. The shaded bands represent the exponents $\zeta^\parallel_p$ defined in 
\S\ref{mixed_sf} obtained from mixed structure functions.

Further analysis indicates that the dissipation field is well-approximated by lognormal statistics. 
The top right panel of figure~\ref{fig:RSS} shows the probability density function of $\log \varepsilon_\ell$, standardised to zero mean and unit variance, for scales $0.2\ell_\nu \leq \ell \leq 10\ell_\nu$. While statistics for small dissipation (left branches) barely superpose, those for strong dissipative events largely collapse on a master curve, with only slight deviations from a Gaussian. 
The bottom right panel of figure~\ref{fig:RSS} shows that the dependency of the mean and variance of $\log \varepsilon_\ell$ with the coarse-graining scale $\ell$, revealing a scaling behaviour reasonably captured by   
\be
	\langle \log(\varepsilon_\ell/\varepsilon) \rangle  = \mu_{62}\log(\ell/\ell_\nu), \quad \mathrm{Var}\,[\log(\varepsilon_\ell/\varepsilon)] = -2 \mu_{62}\log(\ell/\ell_\nu).
\ee
with $\mu_{62} = 0.16 \pm0.02$. Apart from the specific value of the constant $\mu_{62}$, these observations are akin to those commonly observed in 3D homogeneous isotropic turbulence \citep{iyer2015refined,chevillard2019skewed}. They suggest that the scaling exponents of the dissipation moments behave as $\chi_q \approx \mu_{62}\,q\,(1-q)$.

\subsubsection{Nonlinarity of the scaling exponents}
\begin{figure}
\centering
    \includegraphics[width=0.59\linewidth]{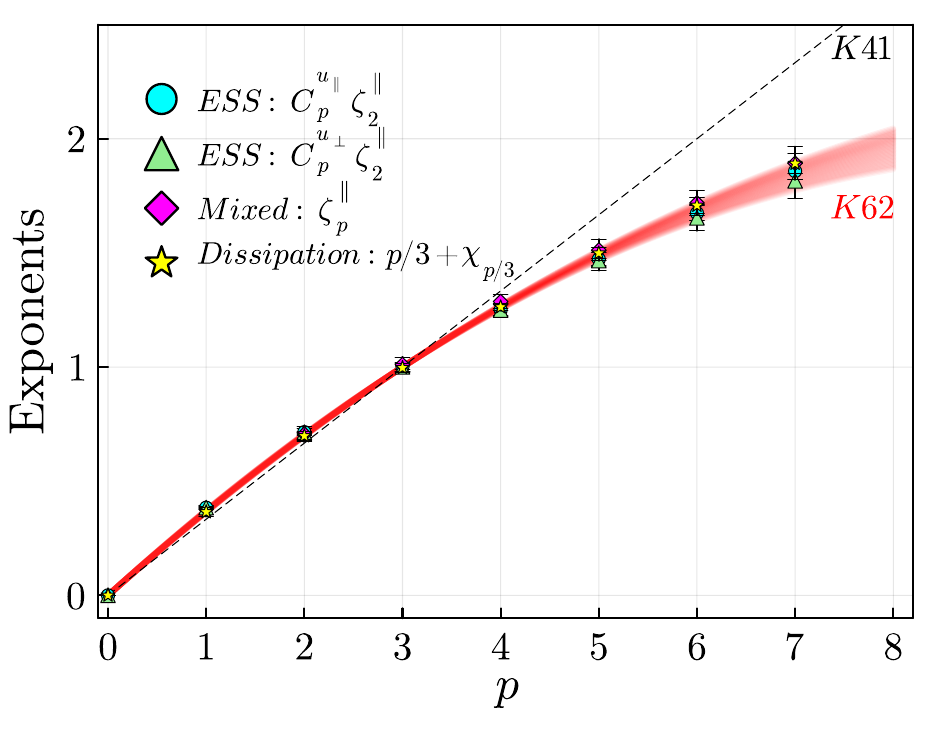}
  \caption{Scaling exponents extracted from different methods: Extended Self Similarity (ESS: $\delta u_\parallel$\textcolor{cyan}{$\bullet$}, $\delta u_\perp$\textcolor{green}{$\blacktriangle$}), mixed structure functions (\textcolor{magenta}{$\mathbin{\blacklozenge}$}), and coarse-grained dissipation (\textcolor{yellow}{$\star$}).  Error bars indicate the root-mean-squared errors.  The dashed black line shows K41 scaling  and the red shaded area represents the log-normal models $\zeta_p = p/3 +\mu_{62}(p/3)(1-p/3)$, with $\mu_{62} \in [0.14,0.18]$.
    }
  \label{fig:intermittency}
\end{figure}
The three types of scaling exponents (mixed, ESS, refined) prove to be remarkably consistent between them. Figure~\ref{fig:intermittency} shows their behavior as function of the oder $p$. For ESS, where the value $ C_p^\phi$ represents a ratio of exponents,  we report the value of  $\zeta^\parallel_2 C_p^\phi$, with $\zeta_2^\parallel$  obtained the mixed exponent of order 2. For all measurements, error bars represent the standard deviation of the local exponent within the inertial range.

In spite of the lack of usual scaling range, figure~\ref{fig:intermittency} suggests that SQG intermittency is characterised by nonlinear scaling exponents  deviating from K41 similarity. The log-normal modelling of the dissipation field provides a reasonable fit up to orders $p\sim 6-7$ with $\mu_{62} \simeq 0.16\pm 0.02$. 
 We observe no sign of saturation for the $\zeta_p$'s  at least up to the order $p=7$ that we computed. 
We however point out that in Navier--Stokes simulations (2D or 3D)  \citep{muller2025lack,iyer2020scaling},   and passive scalar studies \citep{celani2000universality,celani2001fronts,iyer2018steep}, asymptotic behaviours 
of the exponents become apparent only at higher orders $p\gtrsim 10$.

Finally, the correspondence between dissipation and mixed structure function exponents suggests that instead of equation \eqref{eq:bridge}, a consistent refined K62 phenomenology in SQG is mediated through
\be
	\label{eq:bridgesqg}
	 \av{\delta u_\parallel |\delta \theta|^{p-1}} \sim \ell^{p/3} \langle\varepsilon_\ell^{p/3}\rangle,
\ee
entailing the identification $\zeta_p^\parallel = \xi_p=\chi_{p/3}(\ell) + p/3$ between the mixed and refined scaling exponents. 

Note that equation \eqref{eq:bridgesqg} features the fluxes associated to the Casimirs $\av{\theta^{p-1}}$ in its lhs. This, maybe, is counterintuitive. A local version of Yaglom's law in the spirit of  \cite{duchon2000inertial} might suggest the identification $ \delta u \delta \theta^2 \equiv -2\varepsilon_\ell \ell $ entailing rather $S_p^{\rm WG} \sim \langle\varepsilon_\ell^{p/3}\rangle \ell^{p/3}$ instead of \eqref{eq:bridgesqg}. This naive guess  is however not correct:   \S\ref{mixed_sf} has ruled out scaling behaviors for the Watanabe-Gotoh
structure functions.


\section{Concluding remarks}
\label{section5}
In this paper, we investigated forced-dissipated SQG turbulence in a regime dominated by the direct cascade of scalar variance. The aim was to refine the analogy between SQG and Navier--Stokes turbulence, grounded in the Kolmogorov similarity prediction that $|\delta \bu| \sim |\delta \theta| \sim \ell^{1/3}$. Using systematic numerical simulations with up to $16,384^2$ collocation points, we confirm that the two systems share %
 many 
key turbulent features, including anomalous dissipation, exact laws for third-order structure functions, a skewness phenomenon, and intermittency in the form of anomalous scaling. This substantiates the early observations of 
\cite{sukhatme2002surface} in an unforced setting, which notably show that SQG flows are strongly intermittent and, in that sense, closer to the 3D direct cascade than to the 2D inverse cascade. 

However, %
 the idea that $\theta$ is 
 a scalar analogue of a 3D turbulent velocity needs to be promoted only with great care. The analogy stands at the price of considering mixed statistics, involving correlations between the velocity field and the scalar field. In particular, our analysis  points towards the lack of scaling range in the usual sense for structure functions associated to SQG fields $\theta$, $u_\parallel$, $u_\perp$, alone.  This specificity does not fit into usual frameworks of Navier--Stokes turbulence. 
The discrepancy between single-field and mixed statistics is revealed in particular by the specific form of the skewness phenomenon: While Yaglom's law prescribes linear inertial-range scaling for the mixed structure function $\langle \delta \bu \delta \theta^2\rangle = -2 \varepsilon {\boldsymbol \ell} $, it does not prescribe any of the skewnesses for single fields. The increments of scalar and transverse velocities are symmetric and the longitudinal  velocity increments are positively skewed. This is unlike 3D direct cascade, where longitudinal increments are negatively skewed due to the 4/5 law. In SQG, the emerging positive skewness of $\delta u_\parallel$ does not stem from Yaglom's law and instead relates to the balance between subleading Yaglom's terms and the ageostrophic component.

In fact, scaling laws in SQG are structured by the fluxes and anomalous dissipation of the infinite set of Casimir invariants, leading to power-law behaviours for the mixed structure functions $\av{\delta u_\parallel |\delta \theta|^{p-1}}$.  
At the level of single-field statistics, the combination of non-locality (through the Riesz transform),  non-trivial correlations, and geometry blur the Eulerian description.  Still, multi-scaling properties can be quantified through kinematic scaling exponents, obtained from extended and refined similarity analysis. The exponents are all consistent, pointing towards an SQG formulation of the refined-similarity 
 phenomenology
as $ \av{\delta u_\parallel |\delta \theta|^{p-1}} \propto \ell^{p/3} \langle\varepsilon_\ell^{p/3}\rangle$. 

Naturally, one open question relates to the universality of our results. Due to the presence of a large-scale damping in our forcing protocol, only a fraction of the energy input leaks towards the small scales. It is reasonable to expect that various phases of SQG statistics could be identified upon increasing or decreasing this leakage.
Another question relates to the connection with Lagrangian transport. As an active scalar, the Eulerian scaling properties (or lack thereof) in SQG connect to Lagrangian dispersion of particle clusters. Exploring this connection, which extends to the larger issue  of Lagrangian and Eulerian predictability  is a matter for future works.

\backsection[Acknowledgements]{The authors acknowledge stimulating discussions with G.\ Boffetta, C.~Campolina, S.~Musacchio, \'{E}.\ Simonnet, V.\ Valad\~ao. We also thank A. Mailybaev and the Instituto de Matemática Pura e Aplicada, Rio de Janeiro, for support and hospitality (through summer visiting grants), where a part of this work was undertaken. 
The authors are grateful to the Université Côte d’Azur’s Center for High-Performance Computing (OPAL infrastructure) for providing resources and support.
}

\backsection[Fundings]{This work was supported by the French National Research Agency (ANR project TILT; ANR-20-CE30-0035) and the French government through the France 2030 investment plan managed by the ANR, as part of the Initiative of Excellence Université Côte d’Azur under reference number ANR-15-IDEX-01.}

\backsection[Declaration of interests]{The authors report no conflict of interest.}

\backsection[Data availability statement]{The data that support the findings of this study are available upon reasonable request.}

\appendix
	\section{Standard features of SQG}
	\label{sec:SQGfeatures}

	\subsection{Prototype SQG flows}
	\label{ssec:SQGprototypes}
\begin{figure}
	\includegraphics[width=\textwidth]{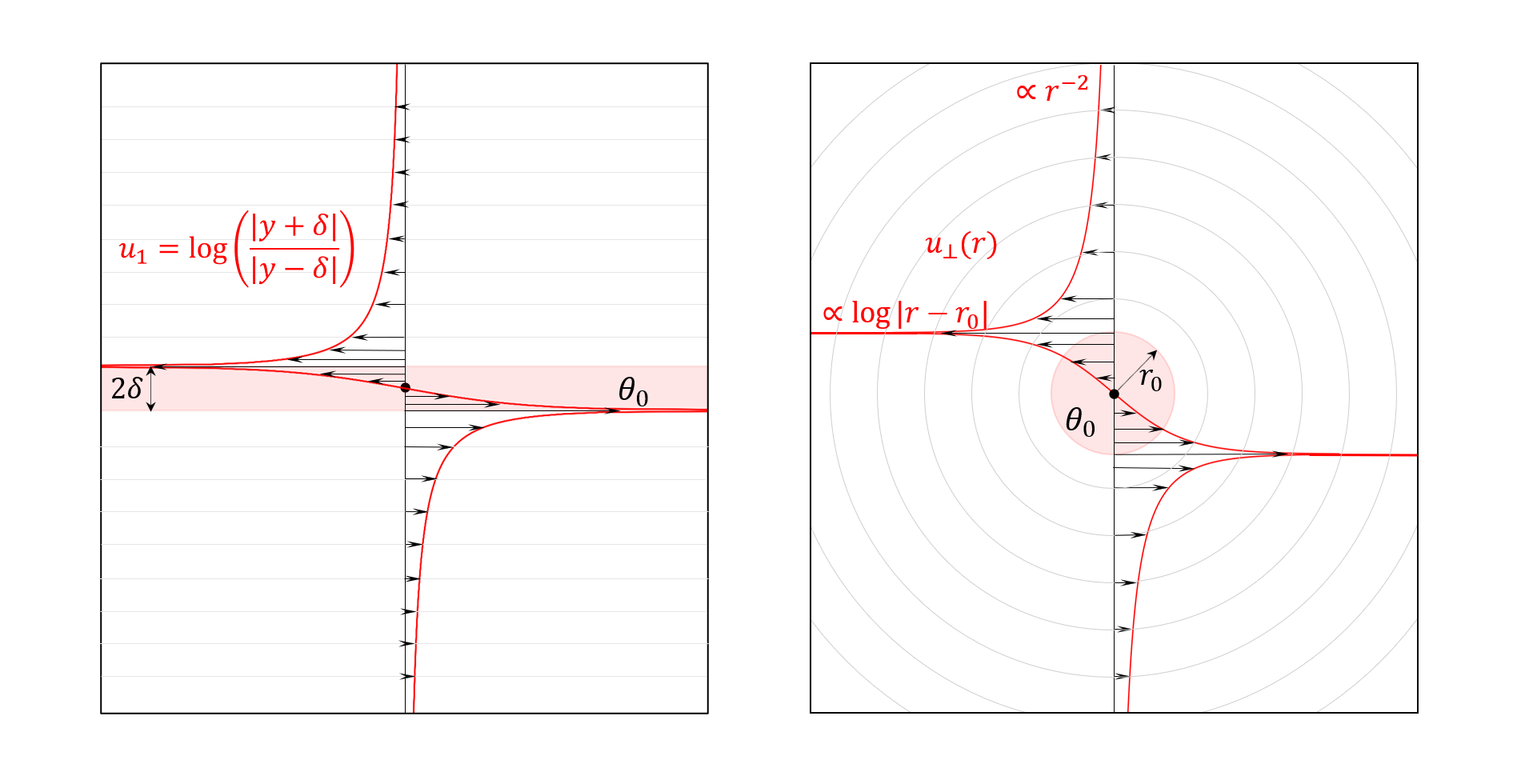}
	\caption{Filaments (left) and vortices (right) as two prototypes of inviscid SQG flows. }
	\label{fig:vortexandfilaments}
	\end{figure}

	It is useful to have in mind prototype inviscid SQG flows, in order to depict SQG turbulence as a mixture of filaments and vortices.
	We refer the reader to \cite{juckes1995instability} for the relevant stability analysis and to figure~\ref{fig:vortexandfilaments} for a sketch of those specific SQG flows in the inviscid case.
	
	\subsubsection*{Filaments}
	SQG filaments  of width $2\delta$, say  aligned along $x_1$,   are prescribed by the scalar field
	\be
	\theta(\bx) = \left\{
		\begin{array}{ll}
			\dfrac{U_0}{2} &, \mbox{ } |x_2|\leq \delta  \\
			0 & \mbox{otherwise}
		\end{array}
	\right.,
	\ee
	which in turn generate a horizontal velocity   $\bu (\bx) = U_0\log\left| \frac{x_2 - \delta}{x_2+\delta}\right| {\hat{\bf 1}}$.
	SQG filaments prescribe two counter propogating jets,  diverging logarithmically at the  boundaries of the filament.  Far from the filament, the horizontal velocity decreases as $u_1(\boldsymbol{x}) \underset{}{\sim} -2U_0\delta x_2^{-1}.$ 
	\subsubsection*{Rankine patches}
	In analogy to the 2D Rankine vortex,  Rankine SQG patches of radius $r_0$ are prescribed by the scalar field
	
	\be
	\theta(\bx) = \left\{
		\begin{array}{ll}
			\theta_0, & \mbox{} |\boldsymbol{x}|\leq r_0  \\
			0 & \mbox{otherwise}
		\end{array}
	\right.,
	\ee
	and prescribe the circular flow
	\be
		\bu(\bx) = u_\perp \hat \bx^\perp,\quad u_\perp:= \theta_0   \int^{+\infty}_{0} \mathcal{J}_1\left(v \frac{\bx}{r_0}\right)\mathcal{J}_1\left(v\right) \, dv
	\ee
	where $\mathcal{J}_{0}$ and $\mathcal{J}_{1}$ are Bessel functions of the first kind, with order $0$ and $1$. 
	At the boundary of the patch, the flow has a logarithmic divergence 
	and far from the patch,  the  velocity decays algebraically, namely
	\be
	u_\perp (\bx) \underset{r_0^\pm}{\propto}  \log\left| r_0-|\bx| \,\right|,\quad u_\perp(\bx)  \underset{+\infty}{\propto} |\boldsymbol{x}|^{-2}.
	\ee
	
	\subsection{Symmetries}
	\subsubsection{Classical symmetries}
	\label{ssec:classical}
	Similar to NS equations, the  SQG system \eqref{eq:sqg} considered in $\mathbb R^2$ is  formally invariant  under a certain number of continuous symmetries. Following the formulations of \cite{frisch1995turbulence}, we list below the symmetries holding in particular in the inviscid case: 
	\begin{itemize}
	\item \emph{Time  or space translation:} $t\mapsto t + \tau,\quad \bx \mapsto \bx + \br$ with $\tau \in \mathbb{R}$ and $\br \in \mathbb{R}^2$.
	
	\item \emph{Rotation and reflection:} $\bx \mapsto \mathcal O\bx,\quad  \bu \mapsto \mathcal O \bu,\quad \theta \mapsto det(\mathcal{O})\theta,$
	with $\mathcal O \in O(2)$.  Pure rotations correspond to $\mathcal O \in SO(2)$, characterized by $\det \mathcal O=1$. Reflections correspond to $\mathcal O \in O(2) \backslash SO(2)$, with $\det \mathcal O=-1$.
	
	\item \emph{Time or space rescaling:} $t\mapsto \alpha t,\quad \bx \mapsto \lambda \bx,\quad\bu \mapsto (\lambda/\alpha) \bu,\quad \theta \mapsto(\lambda/\alpha) \theta$ with $\alpha, \lambda >0$.
	
	\item  \emph{Galilean invariance:}
			$\theta \mapsto \theta + \theta_0,\quad \bu \mapsto \bu + \bu_0,  \quad  \bx \mapsto \bx + t\bu_0$ with $\theta_0 \in \mathbb{R}, \bu_0 \in \mathbb{R}^2$.
	Please note, that to be properly formulated the Galilean invariance needs to be associated  to suitable transformed Green functions \citep{jackson2021classical}, ensuring the non-vanishing asymptotics $\bu  \sim \bu_0$, $\psi \sim - \bx^\perp \cdot \bu_0$  as $|\bx| \to \infty$. This proper formulation however goes beyond  the scope of this work. 
	\end{itemize}
	
	Taking for example  space-time rescaling, the formulation specically means  that 
	if the pair $\bu(\bx,t),\theta(\bx,t)$ solves the inviscid SQG system, then so does the pair
	\be
		\bu'(\bx',t') := (\lambda/\alpha)\bu(\bx'/\lambda,t'/\alpha),\quad \theta'(\bx',t'):=(\lambda/\alpha)\theta'(\bx'/\lambda,t'/\alpha).
	\ee

	\subsubsection{Statistical symmetries}
	\label{sym_cons}
	The statistical symmetries  are statistical formulation of the classical symmetries, which hold at the level of the turbulent averages $\av{.}$.  They mean 
	\be
		\bu'(\bx',t') , \theta'(\bx',t') \overset{law}{\sim} \bu(\bx,t) , \theta(\bx,t),
	\ee
	where the transforms  $\bx,t,\bu,\theta \mapsto \bx',t',\bu',\theta'$ are specified by the classical symmetries listed in \S\ref{ssec:classical}, and respectively yielding stationnarity or homogeneity, isotropy or mirror symmetry, and scale invariance. We let aside the Galilean invariance, which is not discussed in the present paper.
	
	By construction, the statistical symmetries have consequences for the distributions of the SQG fields. We list below the ones with relevance for the present work.
    \begin{itemize}
		 \item One-point statistics:
		\be
			p_\theta(y)  = p_\theta(-y),\quad p_{u_i}(y)  = p_{u_i}(-y),
		\ee
	where $p_\theta(y)= \av{\delta(\theta(\bx_0)-y)}$ is the one-point PDF of $\theta$, and   $p_{u_1}(y)$, $p_{u_2}(y_0)$ that of  each velocity component, with according definitions. For $\theta$,  the symmetry stems from the identities
	$\av{e^{i\theta(\bx_0)}}=\av{e^{i\theta(0)}} = \av{e^{-i\theta(0)}}$, employing successively homogeneity and mirror symmetry, and implying that all odd moments vanish. For the velocity, the same computation holds true, from isotropy instead of mirror symmetry.
	\item Increment statistics:	
		\be
			p_{\delta\theta}(y)  = p_{\delta\theta}(-y),\quad p_{\delta u_\perp}(y)  = p_{\delta u_\perp}(-y),
		\ee
		implying that the odd order moments vanish: $\av{\delta \theta^{2p+1}}=\av{\delta u_{\perp}^{2p+1}}=0$.
	
		\item Mixed structure functions:
		\be
		\forall p \in \mathbb N\, \av{\delta u_{\parallel}\delta\theta^{2p+1}} =\av{\delta u_{\parallel}\delta u_\perp^{2p+1}}=0,\quad \av{\bu(\bx)\theta^2}=0,\quad \av{\bu(\bx) \omega^2}=0 
	\ee
	The second identity stems from the observations $\nabla  \cdot \av{\bu(\bx)\theta(0)^2}=0$ (incompressibility) and $\nabla^\perp  \cdot \av{\bu(\bx)\theta^2}=\av{\omega(\bx) \theta^2} = 0$ (mirror symmetry). Similar arguments apply for the third identity.
	\end{itemize}

	\section{Derivations of Bernard's balances}
	\label{appendix:bernard}
	
	This section summarizes the technical derivation leading to the SQG versions of Bernard's  balances of \S \ref{ssec:kinematicskewness}, either in their complex or differential formulations. 
	
	\subsection{The complex formulation \eqref{complexBernard}}
	Following standard conventions in complex analysis (see, \emph{e.g.}, \cite{tao2016complex}), we introduce the (Wirtinger) complex derivatives as
	$\partial_z = \frac{1}{2} (\partial_1 -i\partial_2)$ and $\partial_{z^*} = \frac{1}{2} (\partial_1 +i\partial_2)$.
	Combining the velocity components into  a single  complex velocity field $U(z,z_*)= u_1+ i u_2$, and  promoting the vorticity and the stream function into the  complex fields  $ \Omega(z,z^*), \Psi(z,z^*)$, one checks the relations
	\be
		\label{eq:complexrelations}
		U(z,z*)   = -2i \partial_{z^*} \Psi,\quad \Omega = -2 i \partial_z U,\quad \Omega=-4 \partial_{zz^*} \Psi,
	\ee
	where  the second identity uses incompressibility.
	We now define the structure functions
	\begin{equation}
	\label{eq:S3s}
		S_3^U(\xi,\xi^\star) = \left\langle  \delta U^3  \right\rangle,\quad   S_3^{\Theta}(\xi,\xi^\star)=\left\langle \left(\delta \Theta\right)^2 \delta U  \right\rangle,\quad S_3^{\Omega}(\xi,\xi^\star)=\left\langle \left(\delta \Omega\right)^2 \delta U  \right\rangle,
	\end{equation} 
	where  $\delta F = F(z+\xi,z^*+\xi^*)-F(z,z_*)$ denotes  complex increments of the field $F$.
	
	With the shorthands $U=U(z,z^*)$, $U'=U(z+\xi,z^*+\xi^*)$, etc,   routine calculations employing statistical homogeneity, isotropy and mirror symmetry yield 
	\be
		\label{eq:S3hom}
		S_3^U= 6\av{U^2U'},\quad S_3^\Theta= 4\,\av{U \Theta \Theta'},\quad S_3^\Omega= 4\,\av{U \Omega \Omega'};
	\ee
	Let us simply emphasize the crucial role played by  the mirror symmetry, which makes  the correlators  $\av{U'\Omega^2}, \av{U'\Theta^2}$  vanish!
	From Eq.~\eqref{eq:S3s} and homogeneity,  the following calculation string holds:
	\be
	\partial_{\xi\xi}\av{(\delta U)^3} = -6\av{(\partial_z'U')\partial_z(U^2)} = 3\av{U\Omega \Omega'},
	\ee
	Using Eq.~\eqref{eq:S3hom}, we then recover the kinematic relation derived  in \cite{bernard1999three}, namely
	\be
		\label{eq:bernardwU}
		\partial_{\xi\xi}S_3^U   = \dfrac{3}{4} S_3^\Omega. 
	\ee
	Being of purely kinematic nature, the relation \eqref{eq:bernardwU} holds true both in SQG and in 2D Navier-Stokes. 
	In order to obtain the SQG balance \eqref{complexBernard}, we  now introduce the ageostrophic field and his complex counterpart as, respectively,
	\be
		\ba = (-\Delta)^{1/2}\left(\bu\theta\right)-\bu\omega,\quad  A:= a_1+ia_2= 2\left|\partial_{z}\right|[U\Theta]-U\Omega,
	\ee
	where the shorthand notation  $|\partial_{z}|$ stands for  the complex counterpart of $\frac{1}{2}(-\Delta)^{1/2}$, such that $\Omega=2|\partial_{z}| \Theta $ and $|\partial_{z}|^2 = -\partial_{zz*}$.

	  We then observe
	\be
		\label{eq:string2}
		\partial_{\xi\xi^*}\av{U\Theta\Theta'} = \av{|\partial_{z}|\left[ U \Theta\right]|\partial_{z'}|\left[\Theta'\right]} =  \dfrac{1}{4} \av{\left(A+ U\Omega\right) \Omega'},
	\ee
	pointing out that the lack of a minus sign in the first equality comes from homogeneity.
	Combining Eq.~\eqref{eq:string2} and  \eqref{eq:S3hom} lead to
	\be
		\label{eq:Theta-Om}
		\partial_{\xi\xi^*}S_3^\Theta = \av{A \Omega'}+\dfrac{1}{4} S_3^\Omega.
	\ee
	Combining Eq.~\eqref{eq:bernardwU} and Eq.~\eqref{eq:Theta-Om} finally yields 
	\be
		\label{complexBernard2}
		\partial_{\xi\xi^*}S_3^\Theta = \av{A \Omega'}+\dfrac{1}{3} \partial_{\xi\xi}S_3^U,
	\ee
	which is the desired formula \eqref{complexBernard}. 
	
	\subsection{The differential formulation \eqref{eq:bernardbalance}}
	Deriving Eq.~\eqref{eq:bernardbalance}  from Eq.~\eqref{complexBernard} relies on algebraic manipulations and the theory of isotropic tensors.
	\subsubsection{Isotropic tensors in dimension $d$}
	 Let us first recall general relations on isotropic tensors. We follow \cite{brachet2000primer}, to observe that the the third-order-but-one-separation isotropic tensors 
	\begin{equation}
		B_{ijk}({\bl}) =\left\langle \delta u_i \delta u_j \delta u_k \right\rangle,\quad \delta u_i:= u_i({\bl})-u_i(0)
	\end{equation}
	can be written as
	\begin{equation}
		\label{eq:BrachetParam}
		B_{ijk}({\bl}) = 2 \left(b_{ij,k}+b_{ki,j}+b_{jk,i}\right),  \quad b_{ij,k} := \left\langle u_i(0)u_j(0)u_k({\bl})\right\rangle, 
	\end{equation}
	where homogeneity obviously implies the identity
	$b_{ij,k} = -\av{u_i({\bl})u_j({\bl})u_k(0)}$.  As an isotropic tensor,  $b_{ij, k}$ takes the form 
	\begin{equation}
	b_{ij, k} = C(\ell)\dfrac{\delta_{ij}\ell_k}{\ell}+D(\ell)\dfrac{\delta_{ik} \ell_j+\delta_{jk} \ell_i}{\ell} +F(\ell)\dfrac{\ell_i\ell_j\ell_k}{\ell^3}.
	\end{equation}
	with the incompressibility condition $\partial_k b_{ij,k}=0$ imposes
	\begin{equation}
	\label{eq:incompressibility}
	-2D = (d-1)C+rC'\quad\&\quad 
	F= -C + \ell \partial_\ell C,
	\end{equation}
	with $d$ denoting space dimensionality ($d=2$ for SQG);  This leaves the function $C$ as the only unknown.
	Direct calculation lead to the following useful  properties:
	\begin{subequations}
	\begin{align}
	& \left\langle \delta u_\parallel^3 \right\rangle  = B_{ijk} \hat \ell_i\hat \ell_j\hat \ell_k = 6(1-d)C \label{eq:brachetpropertiesa}\\
	& \left\langle \delta u_\parallel |{\delta \bf u}|^2 \right\rangle = \hat \ell_iB_{ijj} =2(1-d)\left((1+d)C+\ell\partial_\ell C\right)\label{eq:brachetpropertiesb}\\
	& \left\langle \delta u_\perp |{\delta \bf u}|^2 \right\rangle =  \hat n_iB_{ijj}=0\label{eq:brachetpropertiesc}
	\end{align}
	\end{subequations}
	
	\subsubsection{SQG}
	We now focus on  the case $d=2$, relevant for SQG.   
	We first observe that the complex structure functions $S^U,S^\Theta$ defined in \eqref{eq:S3s} relate to their real counterpart as 
	\be
		\label{eq:S3-realcomplex1}
		\begin{split}
		& \ell S^U = \left(\av{\|\delta u\|^2\delta u_\parallel}+ i\av{\|\delta u\|^2\delta u_\perp}\right)\xi ,\quad 
		 \ell S^\Theta = \left(\av{|\delta \theta|^2\delta u_\parallel}+ i\av{|\delta \theta |^2\delta u_\perp}\right)\xi.
		\end{split}
	\ee
	In line with Eq.~\eqref{eq:brachetpropertiesa}, isotropy and mirror symmetry make the perpendicular contributions vanish,
	such that the previous equation in fact reduces to
	\be
		\label{eq:S3-realcomplex2}
		\begin{split}
		& \ell S^U =  \av{|\delta u|^2\delta u_\parallel}\xi ,\quad 
		 \ell S^\Theta = \av{|\delta \theta|^2\delta u_\parallel} \xi.
		\end{split}
	\ee

	On the one hand,   now combining  \eqref{eq:incompressibility} and
	 \eqref{eq:brachetpropertiesa}, we obtain
	\begin{equation}
	   \label{eq:sqg-b1}
	   C(\ell) = -\dfrac{1}{6}\left\langle \delta u_\parallel^3\right\rangle,\quad  S_3^U = 6\dfrac{F}{\ell^3}\xi^3, \quad F:= \ell^{2}\partial_\ell \left(\ell^{-1}C \right).
	\end{equation}
	Using $\partial_\xi f(\ell ) = \xi^*\partial_\ell f/(2\ell)  $, this in turn yields the first piece of the differential formulation:
	\begin{equation}
	   \label{eq:sqg-b2}
		\begin{split}
		  & \partial_{\xi\xi}S_3^U  = \dfrac{3\xi}{2\ell^3}\partial_\ell\left(\ell^{-1} \partial_\ell \ell^3 F \right)=3\dfrac{\xi}{\ell} \mathcal L_{u_\parallel}\left[\av{\delta u_\parallel^3}\right],\\
		 \text{identifying } &\mathcal{L}_{u_\parallel} := -\frac{1}{12}\ell^{-2}\partial_{\ell} \left(\ell^{-1} \partial_{\ell}\left(\ell^5\partial_{\ell}\left( \ell^{-1} \cdot \right)\right)\right).
		\end{split}
	\end{equation}
	On the other hand, we also compute  from Eq.~\eqref{eq:S3-realcomplex2}, using $\partial_{\xi^*} f(\ell ) = \xi\partial_\ell f/(2\ell)  $:
	\be
	   \label{eq:sqg-b3}
		\partial_{\xi\xi^*} S_3^\Theta = \dfrac{\xi}{\ell} \mathcal L_M\left[\av{|\delta \theta|^2\delta u_\parallel}\right],\quad \mathcal{L}_{M}:= \dfrac{1}{4}\partial_\ell \left(\ell^{-1}\partial_\ell \left( \ell \cdot\right)\right).
	 \ee
	Similar to Eq~\eqref{eq:S3-realcomplex1}, we also have
	\be
	\label{eq:S3-realcomplex3}
		\av{A\Omega'} = -\av{A'\Omega}=-\frac{\xi}{\ell}\left( \av{a_\parallel(\ell {\bf e_1})\, \omega(0)}+i\av{a_\perp(\ell {\bf e_1})\, \omega(0)}\right),
	\ee
	Using    Eq.~\eqref{complexBernard2} with Eq.~\eqref{eq:sqg-b2}, \eqref{eq:sqg-b3} and \eqref{eq:S3-realcomplex3} yields
	$\av{a_\perp(\ell {\bf e_1})\,\omega(0)}=0$ and the differential formulation  \eqref{eq:bernardbalance}.


\bibliographystyle{jfm}
\bibliography{biblio}

\end{document}